\documentclass[aps,prb,groupedaddress,twocolumn,superscriptaddress,10pt]{revtex4-2}
\usepackage{graphicx}
\usepackage{mathtools}
\usepackage{amsmath}
\usepackage{amssymb}
\usepackage{bbm}
\usepackage{bm}
\usepackage{cancel}
\usepackage{xspace}
\usepackage{braket}
\usepackage{xcolor}
\usepackage{siunitx}
\usepackage{calrsfs}
\usepackage{comment}
\usepackage{placeins}%enables \FloatBarier
\usepackage[export]{adjustbox}
\usepackage{wrapfig}
\usepackage[colorlinks=true]{hyperref} % Use hidelinks to hide highlighting

\usepackage{units}
 
\newcommand{\ee}{\mathrm{e}}  % Euler number
\DeclareMathOperator*{\ii}{i} % imaginary unit
\newcommand*\dd{\mathop{}\!\mathrm{d}}

\renewcommand{\vec}[1]{\bm{#1}} % vectors in bold
\newcommand{\mat}[1]{\bm{#1}} % matrices in bold
\newcommand{\kel}[1]{\underline{#1}} % objects on Keldysh contour

\definecolor{hblue}{RGB}{0,0,255}
\newcommand{\TMch}[1]{{\color{black} #1}}

\newcommand{\resub}[1]{{\color{black} #1}}

\begin{document}

\title{Correlated Mott insulators in a strong electric field: \\ The effects of phonon renormalization}
	
\author{Tommaso Maria Mazzocchi}
\email[]{mazzocchi@tugraz.at}
\affiliation{Institute of Theoretical and Computational Physics, Graz University of Technology, 8010 Graz, Austria}
\author{Daniel Werner}
\affiliation{Institute of Theoretical and Computational Physics, Graz University of Technology, 8010 Graz, Austria}
\author{Paolo Gazzaneo}
\affiliation{Institute of Theoretical and Computational Physics, Graz University of Technology, 8010 Graz, Austria}
\author{Enrico Arrigoni}
\email[]{arrigoni@tugraz.at}
\affiliation{Institute of Theoretical and Computational Physics, Graz University of Technology, 8010 Graz, Austria}

\date{\today}
	
\begin{abstract}

We characterize the response of a Mott insulating system to a static electric field in terms of its conducting and spectral properties.
Dissipation is included by a coupling to fermionic baths and to either optical or acoustic phonons. This paper extends and completes the analysis made in a previous work by the authors [arXiv:2207.01921]. In the present work phonons are included {\em diagrammatically} within the Migdal approximation by also including self-consistency from the electronic feedback. The nonequilibrium steady-state is addressed by means of the dynamical mean-field theory based on the nonequilibrium Green's function approach, while the so-called auxiliary master equation approach is employed as impurity solver. With optical phonons the self-consistency suppresses the steady-state current \resub{for field strengths comparable to the band gap} with respect to the nonself-consistent case. This is due to the interaction of phonons with the hot electrons of the lattice which increases their temperature, thus providing a less effective relaxation channel for the current-induced Joule heat. In addition, in the case of optical phonons the results are essentially independent of the temperature of the fermionic baths, as the latter is sensibly smaller than their characteristic frequency. On the other hand, with acoustic phonons the steady-state current is slightly suppressed by the self-consistent treatment only at field strengths close to half of the gap and especially at very small phonon frequency. Also, in this case the results seem to slightly depend on the temperature of the fermionic baths.
\end{abstract}

%\pacs{}
	
\maketitle

\section{Introduction}\label{sec:intro}

Electrically-driven models of Mott-insulating systems are known to exhibit an insulator-to-metal transition~\cite{aron.12,am.we.12,ec.we.13.db} at large field strengths. This makes them candidates to describe the {\em resistive switch} occurring in Mott insulators and correlated systems~\cite{ja.tr.15} under the action of a constant bias voltage. It is believed that the resistive switch is due to the formation of metallic filaments percolating through the material/device~\cite{ja.tr.15,st.ca.13}. The so-called {\em effective resistors} models~\cite{st.ca.13} or the non-homogeneous mean-field theory~\cite{li.ar.17} are only a couple of possible explanations of this phenomenon, as it is not entirely clear what microscopic mechanism leads to the formation of such filaments.

Far from providing an explanation to the way these filaments are created, the first attempts to model the dielectric breakdown of an insulator have focused on the importance of a fermion bath in the context of dissipative systems~\cite{ts.ok.09} and their role in getting to a non-trivial nonequilibrium steady-state (NESS)~\cite{ar.ko.12,am.we.12}. Also, it is still debated if the resistive switch occurs mainly due to thermal-~\cite{li.ar.15,ha.li.18,di.ha.22u} or quantum-triggered~\cite{ha.ar.22u} effects. However, there is agreement on the fact that a satisfactory understanding of a field-induced dielectric breakdown must take into account the realistic microscopic mechanism leading to Joule heat dissipation. To do so, lattice vibrations, i.e. {\em phonons}, should be included in the model of the insulator together with the electronic degrees of freedom. To this end, a first attempt has been made in~\cite{ha.ar.22u}, focusing on a two-dimensional lattice coupled to either acoustic or Einstein phonons.

A comprehensive description cannot avoid the inclusion of feedback effects onto phonons which may help to characterize the NESS in terms of the possibly temperature-triggered effects onto the dielectric breakdown of the insulator. In fact, due to the large amount of energy required to overcome the insulating phase, \resub{when the electric field equals the band gap} a comparable amount of heat is expected to be generated by the accelerated electrons of the system. In the absence of feedback effects from these hot electrons, the description of the dielectric breakdown would then miss the phonon contributions to the heat transport within the material. For this reason, in this paper we also include self-consistent (SC) phonons on a single-band Hubbard model in a static electric field, as opposed to the nonself-consistent (NSC) case with only acoustic phonons addressed in Ref.~\cite{ma.ga.22}. In addition, we also extend our investigation to the case of an optical phonon branch, which we model as an Einstein phonon coupled to an ohmic bath, in order to assess their effectiveness in dissipating Joule heat. However, as pointed out in~\cite{ma.ga.22}, a non-trivial NESS is quite difficult to reach with phonons alone as dissipation mechanism; for this reason we also couple the system to an electron bath. Our main goal is to characterize the different types of SC phonons -- either optical or acoustic -- in terms of the dissipation of the current-induced Joule heat in a model of a correlated insulator \resub{near the current-conducting state}. \resub{In order to model realistic materials one could take into account for example (i) disorder due to crystal defects, (ii) multi-orbital models, (iii) a spatially inhomogeneous domain structure as presented, e.g., in References~\cite{st.ca.13} and~\cite{li.ar.17}.} For this reason, the description of the electric field-driven insulator-to-metal transition occurring in realistic materials is beyond the purpose of the present paper.

The rest of the paper is organized as follows: In Sec.~\ref{sec:MO_HA} we introduce the model, while in Sec.~\ref{sec:method} we present the Dyson equations for both the electronic and phononic Green's function (GF): we refer to Appendix~\ref{sec:GFs_Dyson_Floquet} for further details concerning the Floquet structure of the latter. In Sec.~\ref{sec:results} we discuss the results and leave Sec.~\ref{sec:conclusions} for final remarks and comments.
	
\section{Model Hamiltonian}\label{sec:MO_HA}
We start from the setup described in Ref.~\cite{ma.ga.22}, namely the single-band Hubbard model in the presence of a constant electric field, the Hamiltonian of which is given by 
\begin{equation}\label{eq:MicroHamiltonian}
\hat{H}(t) = \hat{H}_{\text{U}}(t) + \hat{H}_{\text{bath}} + \hat{H}_{\text{e-ph}} + \hat{H}_{\text{ph}}.
\end{equation}

The Hubbard Hamiltonian $\hat{H}_{\text{U}}(t)$ is given by
\begin{equation}\label{eq:Hubbard_ham}
\hat{H}_{\text{U}}(t) = \varepsilon_{\text{c}} \sum_{i\sigma}\hat{n}^{f}_{i\sigma} -\sum_{\sigma}\sum_{(i,j)} t_{ij}(t) \hat{f}^{\dagger}_{i\sigma} \hat{f}_{j\sigma} + U \sum_{i} \hat{n}^{f}_{i\uparrow} \hat{n}^{f}_{i\downarrow},
\end{equation}
where $\hat{f}^{\dagger}_{i\sigma}$ ($\hat{f}_{i\sigma}$) is the creation (annihilation) operator of an electron of spin $\sigma= \{ \uparrow,\downarrow \}$ at the $i$-th lattice site and $\hat{n}^{f}_{i\sigma}\equiv \hat{f}^{\dagger}_{i\sigma} \hat{f}_{i\sigma}$ the corresponding density operator. Sums over nearest neighbor sites are denoted by $(i,j)$ and the electrons' \emph{onsite energy} is chosen as $\varepsilon_{\text{c}} \equiv -U/2$. 
In the temporal gauge the static homogeneous electric field defines the time dependent hopping $t_{ij}(t)$ in Eq.~\eqref{eq:Hubbard_ham} via the Peierls substitution~\cite{peie.33}
\begin{equation}\label{eq:peierls}
t_{ij}(t) = t_{\text{c}} \ \ee^{-\ii \frac{Q}{\hbar} \left( \vec{r}_j - \vec{r}_i \right) \cdot \vec{A}(t)}, 
\end{equation} 
where $t_{\text{c}}$ is the hopping amplitude, $\vec{A}$(t) the homogeneous vector potential, $Q$ the electron charge and $\hbar$ Planck's constant. The static electric field is then given by $\vec{F}= -\partial_{t}\vec{A}(t)$ where we choose $\vec{A}(t)=\vec{e}_{0} A(t)$, with $\vec{e}_{0}=(1,1,\ldots,1)$ denoting the lattice body diagonal and
\begin{equation}\label{eq:TD_VecPot}
\vec{A}(t)= -\vec{F}\  t.
\end{equation}
By means of Eqs~\eqref{eq:peierls} and \eqref{eq:TD_VecPot} we define the Bloch frequency  $\Omega \equiv -FQa/\hbar$ with $a$ being the lattice spacing and $F\equiv |\vec{F}|$. 
Here we consider a $d$-dimensional lattice in the $d \rightarrow \infty$ limit~\cite{mu.we.18} with the usual rescaling of the hopping $t_{\text{c}}=t^{\ast}/(2\sqrt{d})$. Sums over the crystal momentum are then performed using the joint density of states~\cite{ts.ok.08,ma.ga.22} $\rho(\epsilon,\overline{\epsilon}) = 1/(\pi t^{\ast 2}) \ \exp[-( \epsilon^{2} + \overline{\epsilon}^{2})/t^{\ast 2}]$ with $\epsilon = -2t_{\text{c}} \sum_{i=1}^{d} \cos(k_i a)$ and $\overline{\epsilon} = -2t_{\text{c}}\sum_{i=1}^{d} \sin(k_i a)$.

In this work we {\em attach} either an optical phonon or an acoustic phonon branch to each lattice site. The electron-phonon interaction is given by the Hamiltonian
\begin{equation}\label{eq:e-ph_Ein_ham}
\hat{H}_{\text{e-ph}} = g \sum_{i\sigma} \hat{n}^{f}_{i\sigma} \hat{x}_{i}
\end{equation}
with $\hat{x}_{i}\equiv (\hat{b}^{\dagger}_{i} + \hat{b}_{i})/\sqrt{2}$, where $\hat{b}^{\dagger}_{i}$ ($\hat{b}_{i}$) can either create (annihilate) an optical phonon at the lattice site $i$ or an acoustic phonon belonging to the branch $i$. In the former case, the optical phonon Hamiltonian consists of an Einstein phonon $\hat{H}_{\text{ph},\text{E}} = \omega_{\text{E}}\sum_{i}\hat{n}^{b}_{i}$ with $\hat{n}^{b}_{i}=\hat{b}^{\dagger}_{i}\hat{b}_{i}$ the phonon density, coupled to an ohmic bath $\hat{H}_{\text{ph},\text{ohm}}$ with spectral density given in Eq.~\eqref{eq:ohm_bath_spec}. The details concerning acoustic phonons implementation can be found in our previous work~\cite{ma.ga.22} and in Sec.~\ref{sec:Ph_Dyson} of this paper.

As pointed out in the introduction, a stable steady-state~\footnote{For further details about the stability of the steady-state we point at our recent paper in Ref~\cite{ma.ga.22}.} is hard to reach with phonon-mediated dissipation only. This is due to the fact that the narrow phonon bandwidth~\footnote{Especially when considering optical phonon, the phonon bandwidth is really small with respect to the other energy scales.} cannot relax electrons across the band gap~\cite{ma.ga.22}. For this reason, it is convenient to include fermion baths in the guise of B\"uttiker tube chains attached to each lattice site via the Hamiltonian $\hat{H}_{\text{bath}}$, the details of which will be specified in Sec.~\ref{sec:Dyson-eq}, see Eq.~\eqref{eq:WBL_bathGF}. \resub{We stress that the fermionic baths are introduced as mere theoretical expedients to make the DMFT loop stable: they would not even provide a description of a possible substrate since the latter is insulating.} We set $\hbar = k_{\text{B}} = a = 1 = -Q$, such that the Bloch frequency $\Omega$ equals the electric field strength $F$ and the current is measured in units of $t^{\ast}$. In the following, we denote the electron and phonon GFs by $G$ and $D$, and the corresponding self-energy (SE) by $\Sigma$ or $\Pi$, respectively.

\section{Methods}\label{sec:method}

\subsection{Electron Dyson equation}\label{sec:Dyson-eq}

Here we follow the derivation given in~\cite{ma.ga.22}: for details about the Floquet structure we refer to~\cite{ts.ok.08} and to Appendix~\ref{sec:GFs_Dyson_Floquet}. The Dyson equation for the electronic lattice GF reads
\begin{equation}\label{eq:FullDysonEq}
\kel{\mat{G}}^{-1}(\omega,\epsilon,\overline{\epsilon}) = \kel{\mat{G}}^{-1}_{0}(\omega,\epsilon,\overline{\epsilon}) - \kel{\mat{\Sigma}}(\omega,\epsilon,\overline{\epsilon}) - \kel{\mat{\Sigma}}_{\text{e-ph}}(\omega,\epsilon,\overline{\epsilon}),
\end{equation}
where both electron and e-ph SE depend on the crystal momentum via $\epsilon$, $\overline{\epsilon}$. In this paper, any Floquet-represented matrix is denoted by either $X_{mn}$ or $\mat{X}$ (see e.g.~\cite{so.do.18,ma.ga.22}), while an underline defines the so-called Keldysh structure
\begin{equation}\label{eq:Keld-structure}
\kel{\mat{X}} \equiv 
\begin{pmatrix}
\mat{X}^{\text{R}} & \mat{X}^{\text{K}}\\
\mat{0}         & \mat{X}^{\text{A}} \\
\end{pmatrix}
\end{equation}
with $\mat{X}^{\text{R},\text{A},\text{K}}$ being the {\em retarded}, {\em advanced} and {\em Keldysh} components. We recall that $\mat{X}^{\text{A}}=(\mat{X}^{\text{R}})^{\dagger}$ and $\mat{X}^{\text{K}} \equiv \mat{X}^{>} + \mat{X}^{<}$, where $\mat{X}^{\lessgtr}$ are the \emph{lesser} and \emph{greater} components~\cite{schw.61,keld.65,ra.sm.86,ha.ja}.

The electron GF of the non-interacting part of the Hamiltonian~\eqref{eq:MicroHamiltonian} reads
\begin{align}\label{eq:non-int_InvGF}
\begin{split}
[G_{0}^{-1}(\omega,\epsilon,\bar{\epsilon})]^{\text{R}}_{mn} & = \left[ \omega_n-\varepsilon_c -v^{2}g^{\text{R}}_{\text{bath}}(\omega_n) \right]\delta_{mn} - \varepsilon_{mn}(\epsilon,\overline{\epsilon}), \\
[G_{0}^{-1}(\omega,\epsilon,\bar{\epsilon})]^{\text{K}}_{mn} & = - \delta_{mn} v^{2}g^{\text{K}}_{\text{bath}}(\omega_{n})
\end{split}
\end{align}
with the shorthand notation $\omega_{n}\equiv \omega+n\Omega$. The off-diagonal terms in Eq.~\eqref{eq:non-int_InvGF} are given by the Floquet dispersion relation $\varepsilon_{mn}$ which, for a hypercubic lattice in a dc field~\cite{ts.ok.08}, reads
\begin{equation}\label{eq:Floquet_disp}
\varepsilon_{mn}(\epsilon,\overline{\epsilon}) = \frac{1}{2} \left[ \left( \epsilon + \ii \overline{\epsilon} \right)\delta_{m-n,1} +  \left( \epsilon - \ii \overline{\epsilon} \right)\delta_{m-n,-1} \right].
\end{equation}
We make use of the so-called wide band limit for the electronic bath GF in Eq.~\eqref{eq:non-int_InvGF}, according to which the \emph{retarded} and \emph{Keldysh} components~\cite{ma.ga.22,ne.ar.15} read
\begin{align}\label{eq:WBL_bathGF}
\begin{split}
v^{2}g^{\text{R}}_{\text{bath}}(\omega) & = - \ii \Gamma_{\text{e}}/2, \\
v^{2}g^{\text{K}}_{\text{bath}}(\omega) & = 2\ii \text{Im}[\Sigma^{\text{R}}_{\text{bath}}(\omega)] \tanh \left[ \beta\left(\omega-\mu\right)/2\right]
\end{split}
\end{align}
with $v$ being the hybridization strength between the system and the electron bath, and $\beta$ and $\mu$ the inverse temperature and chemical potential of the bath.

The electron and e-ph SEs $\kel{\mat{\Sigma}}$ and $\kel{\mat{\Sigma}}_{\text{e-ph}}$ are obtained from the dynamical mean-field theory~\cite{me.vo.89,ge.ko.92,ge.ko.96} (DMFT), and its non-equilibrium Floquet (F-DMFT) extension~\cite{ts.ok.08,sc.mo.02u,jo.fr.08}, by means of the approximations $\kel{\mat{\Sigma}}(\omega,\epsilon,\overline{\epsilon}) \approx \kel{\mat{\Sigma}}(\omega)$, $\kel{\mat{\Sigma}}_{\text{e-ph}}(\omega,\epsilon,\overline{\epsilon}) \approx \kel{\mat{\Sigma}}_{\text{e-ph}}(\omega)$. Further details can be found in Appendix~\ref{sec:imp_solver}.

\subsubsection{Electron-phonon SE}\label{sec:e-ph_SE_impl}

Within DMFT, the e-ph SE is taken to be a local quantity too, namely $\kel{\mat{\Sigma}}_{\text{e-ph}}(\omega,\epsilon,\overline{\epsilon}) \approx \kel{\mat{\Sigma}}_{\text{e-ph}}(\omega)$. In terms of the {\em contour-times} $z,z^\prime$, and in the Migdal approximation~\footnote{It should be noted that in the Migdal approximation the so-called {\em Hartree term} amounts to a constant energy shift that can be reabsorbed in a constant factor in the electron-phonon Hamiltonian $\hat{H}_{\text{e-ph}}$ at half-filling.}, the latter reads~\cite{ma.ga.22}
\begin{equation}\label{eq:backbone_e-ph_SE}
\Sigma_{\text{e-ph}}(z,z^{\prime}) = \ii g^{2} G_{\text{loc}}(z,z^{\prime}) D_{\text{ph}}(z,z^{\prime})
\end{equation}
and corresponds to the lowest-order diagram in the phonon propagator $D_{\text{ph}}$, the form of which will be discussed in Secs~\ref{sec:ac_ph_formalism} and~\ref{sec:Ein_ph_formalism}. The {\em retarded} and {\em Keldysh} components of Eq.~\eqref{eq:backbone_e-ph_SE} can be found in Appendix~\ref{sec:real-time_eph_se}. 

Here we only mention that $G_{\text{loc}}(z,z^{\prime})$ is the {\em contour-times} local electron GF allowing the following representation in frequency-domain
\begin{equation}\label{eq:Lat_LocGF}
\begin{split}
\kel{\mat G}_{\text{loc}}(\omega) &= \int \dd\epsilon \int \dd\overline{\epsilon} \ \rho(\epsilon,\overline{\epsilon}) \\ 
&\times \left\{ \left[ \kel{\mat G}^{-1}_{0}(\omega,\epsilon,\overline{\epsilon}) - \kel{\mat\Sigma}(\omega) - \kel{\mat \Sigma}_{\text{e-ph}}(\omega) \right]^{-1} \right\}.
\end{split}
\end{equation}
Due to gauge invariance $\kel{\mat G}_{\text{loc}}(\omega)$ is diagonal in Floquet indices in the case of a dc field~\cite{ts.ok.08,ma.ga.22} considered here. We now separately discuss the setups pertaining acoustic and optical phonons.

\subsection{Phonon Dyson equation}\label{sec:Ph_Dyson}

\subsubsection{Acoustic phonons}\label{sec:ac_ph_formalism}

In this paper we include acoustic phonons by using an {\em ohmic} density of states (DOS)~\cite{pi.li.21,ma.ga.22}

\begin{equation}\label{eq:ac_ohmic_DOS}
\begin{split}
 \rho_{\text{D}}(\omega) & \equiv -\frac{1}{\pi}\text{Im}[D^{\text{R}}_{\text{ph},0}(\omega)] \\
 & = \frac{ \omega}{4\omega^{2}_{\text{D}}} e^{-|\omega|/\omega_{\text{D}}}
\end{split}
\end{equation}

for the unperturbed Hamiltonian $\hat{H}_{\text{ph},0}$. \resub{In our simplified model, acoustic phonons disperse only in a direction perpendicular to the lattice. The reasons are twofold. On the one hand, we are interested in their effect as heat dissipators, so the direction out of the plane is the relevant one. Second, this is somewhat consistent with the DMFT approximation. Due to this geometry, we can integrate out the out-of-plane phonon levels, which leads to the spectrum considered in this Manuscript.} The resulting Dyson equation then reads
\begin{equation}\label{eq:local_Dyson_ac_ph}
\kel{D}_{\text{ph}}(\omega) = [\kel{D}^{-1}_{\text{ph},0}(\omega) - \kel{\Pi}_{\text{e-ph}}(\omega)]^{-1},
\end{equation}
where $\kel{D}^{-1}_{\text{ph},0}(\omega)$ is the non-interacting phonon propagator~\cite{ao.ts.14,pi.li.21}, the real part of which is determined by the Kramers-Kr\"onig relations and the {\em Keldysh} component by the fluctuation-dissipation theorem~\footnote{For bosons the fluctuation-dissipation theorem reads $\Pi^{\text{K}}_{\text{bath}}(\omega) = \left(\Pi^{\text{R}}_{\text{bath}}(\omega) - \Pi^{\text{A}}_{\text{bath}}(\omega)\right) \coth(\beta\omega/2)$.}
\begin{equation}
\label{eq:non-int_acoustic_ph}
D^{\text{K}}_{\text{ph},0}(\omega) = -2\pi \ii \rho_{\text{D}}(\omega) \coth(\beta\omega/2).
\end{equation}

Notice that the ohmic phonon DOS~\eqref{eq:ac_ohmic_DOS} ensures a linear dispersion relation in the low-energy range $\omega \approx 0$.

\subsubsection{Optical phonons}\label{sec:Ein_ph_formalism}

We model the optical phonon branch by Einstein phonons coupled to an ohmic bath, the Dyson equation of which reads
\begin{equation}\label{eq:local_Dyson_Ein_ph}
\kel{D}_{\text{ph}}(\omega) = [\kel{D}^{-1}_{\text{ph},\text{E}}(\omega) - \kel{\Pi}_{\text{bath}}(\omega) - \kel{\Pi}_{\text{e-ph}}(\omega)]^{-1}
\end{equation}
with the non-interacting Einstein phonon propagator 
\begin{align}\label{eq:non-int_einstein_ph}
\begin{split}
D^{\text{R}}_{\text{ph},\text{E}}(\omega) & = 2\omega_{\text{E}}/\left(\omega^{2} - \omega_{\text{E}}^{2}\right), \\
D^{\text{K}}_{\text{ph},\text{E}}(\omega) & \to 0,
\end{split}
\end{align}
in which the Keldysh component can be neglected due to the presence of $\kel{\Pi}_{\text{bath}}$, which will be described below.

The Einstein phonon is coupled to an ohmic bath $\hat{H}_{\text{ph},\text{ohm}}$, the real {\em retarded} GF of which is obtained from the Kramers-Kr\"onig relations (see e.g. Ref.~\cite{mu.ts.17}), while the {\em Keldysh} component is given by
\begin{equation}\label{eq:ohmic_bath_GF}
\Pi^{\text{K}}_{\text{bath}}(\omega) = -2\pi\ii A_{\text{bath}}(\omega) \coth(\beta\omega/2).
\end{equation}
The ohmic bath DOS in~\eqref{eq:ohmic_bath_GF} is taken as
\begin{equation}\label{eq:ohm_bath_spec}
A_{\text{bath}}(\omega) = \frac{v^{2}_{\text{c}}}{\omega_{\text{c}}} \left[ \frac{1}{1+\left( \frac{\omega-\omega_{\text{c}}}{\omega_{\text{c}}}\right)^{2}}  - \frac{1}{1+\left( \frac{\omega+\omega_{\text{c}}}{\omega_{\text{c}}}\right)^{2}} \right]
\end{equation}
with the usual definition $-\pi A_{\text{bath}}(\omega) \equiv \text{Im}[\Pi^{\text{R}}_{\text{bath}}(\omega)]$. In Eq.~\eqref{eq:ohm_bath_spec} $\omega_{\text{c}}$ denotes the ohmic bath cutoff frequency and $v_{\text{c}}$ the hybridization strength to the ohmic bath~\footnote{The parameters $v_{\text{c}}$ and $\omega_{\text{c}}$ are chosen such that $\alpha \ \text{Im}[\Pi^{\text{R}}_{\text{bath}}(\omega_{\text{max}})]<\text{Im}[\Pi^{\text{R}}_{\text{e-ph}}(\omega^{\prime}_{\text{max}})]$ with $\alpha\in [2,3]$ and $\omega_{\text{max}}$, $\omega^{\prime}_{\text{max}}$ being the points at which $\text{Im}\Pi^{\text{R}}_{\text{bath}}$ and $\text{Im}\Pi^{\text{R}}_{\text{e-ph}}$ have their maxima.}. Notice that Eq.~\eqref{eq:ohm_bath_spec} ensures a linear dependence within almost the entire interval $\omega \in [-\omega_{\text{c}},\omega_{\text{c}}]$. 

\subsubsection{Self-consistent phonons and polarization diagram}\label{sec:self-cons_phonons}

According to the DMFT approximation of local SE, the polarization diagram $\Pi_{\text{e-ph}}$ only depends on the local electron GFs. Within the Migdal approximation, the contour times {\em polarization} diagram~\cite{mu.we.15,mu.ts.17} in Eqs~\eqref{eq:local_Dyson_ac_ph} and \eqref{eq:local_Dyson_Ein_ph} reads 
\begin{align}\label{eq:bubble_GG}
\Pi_{\text{e-ph}}(z,z^{\prime})=-2\ii g^{2} G_{\text{loc}}(z,z^{\prime})G_{\text{loc}}(z^{\prime},z)
\end{align}
with $G_{\text{loc}}(z,z^{\prime})$ being the electron GF on the Keldysh contour allowing the representation~\eqref{eq:Lat_LocGF} and the factor $2$ accounting for spin degeneracy. The real time components of Eq.~\eqref{eq:bubble_GG} are also derived in Appendix~\ref{sec:real-time_eph_se}.

\subsection{Observables}\label{sec:observables}

The local electron and phonon spectral functions read
\begin{align}\label{eq:local_spec_func}
\begin{split}
A(\omega)& =-\text{Im}[G^{\text{R}}_{\text{loc}}(\omega)]/\pi, \\ 
A_{\text{ph}}(\omega)& =- \text{Im}[D^{\text{R}}_{\text{ph}}(\omega)]/\pi.
\end{split}
\end{align}
We define the electron spectral occupation function as
\begin{equation}\label{eq:Filling_func}
N_{\text{e}}(\omega) \equiv A(\omega)\left\{ \frac{1}{2} - \frac{1}{4} \frac{\text{Im}[G^{\text{K}}_{\text{loc}}(\omega)]}{\text{Im}[G^{\text{R}}_{\text{loc}}(\omega)]} \right\},
\end{equation}
where the combination in curly brackets is the nonequilibrium electron distribution function
\begin{equation}\label{eq:NEFD-dist}
F_{\text{el}}(\omega) \equiv \frac{1}{2} \left\{1 - \frac{1}{2}\frac{\text{Im}[G^{\text{K}}_{\text{loc}}(\omega)]}{\text{Im}[G^{\text{R}}_{\text{loc}}(\omega)]} \right\}.
\end{equation}
Analogously, we define the nonequilibrium phonon distribution function $F_{\text{ph}}(\omega)$ as 
\begin{equation}\label{eq:NEBD-dist}
F_{\text{ph}}(\omega) = -\frac{1}{2} \left\{1 - \frac{1}{2}\frac{\text{Im}[D^{\text{K}}_{\text{ph}}(\omega)]}{\text{Im}[D^{\text{R}}_{\text{ph}}(\omega)]} \right\}.
\end{equation}
In our units, the steady-state current~\cite{ma.ga.22} reads~\footnote{We recall that due to the time-independent nature of the dc field setup~\cite{ts.ok.08,ma.ga.22} the elements with $l\neq 0$ of any Wigner-represented matrix are vanishing.}
\begin{equation}\label{eq:general_Wig_current}
\begin{split}
J = \int_{-\infty}^{+\infty} & \frac{\dd\omega}{2\pi} \int \dd\epsilon \int \dd\overline{\epsilon} \ \rho(\epsilon,\overline{\epsilon}) \\
& \times \left[ \left( \epsilon - \ii \overline{\epsilon} \right) G^{<}_{1}(\omega,\epsilon,\overline{\epsilon}) + \text{H.c.} \right],
\end{split}
\end{equation}
while the steady-state kinetic energy is given by
\begin{equation}\label{eq:general_Wig_energy}
\begin{split}
E_{\text{kin}} = \int_{-\infty}^{+\infty} & \frac{\dd\omega}{2\pi} \int \dd\epsilon \int \dd\overline{\epsilon} \ \rho(\epsilon,\overline{\epsilon}) \\
& \times \left[ - \left( \overline{\epsilon} + \ii\epsilon \right) G^{<}_{1}(\omega,\epsilon,\overline{\epsilon}) + \text{H.c.} \right].
\end{split} 
\end{equation}

\begin{table}[b]
  \begin{center}
\begin{tabular}{ cccccccccc }
      \hline
      \hline
      & $U/t^{\ast}$ & $\varepsilon_{\text{c}}/t^{\ast}$ & $\mu/t^{\ast}$ & $\beta/t^{\ast-1}$ & $\Gamma_{\text{ph}}/t^{\ast}$ & $\omega_{\text{c}}/t^{\ast}$ & $g/t^{\ast}$ & $\omega_{\text{E}}/t^{\ast}$ & $\omega_{\text{D}}/t^{\ast}$ \\
      \hline
     O & 8 & -4 & 0 & 20 & 23.8   & 0.6  & 0.4 & 0.6  & 0   \\
     A & 8 & -4 & 0 & 20 & 1.850  & 0    & 0.4 & 0    & 0.05 \\
      \hline
      \hline
    \end{tabular}
    \caption{Default parameters for electron bath plus optical (setup O) and acoustic (setup A) phonons. In setup O the phonon coupling strength is defined as $\Gamma_{\text{ph}}\equiv 2\pi g^{2} \rho_{\text{E}}(\omega_{\text{E}})|_{\text{NSC}}$, where $\rho_{\text{E}}(\omega_{\text{E}})|_{\text{NSC}} = -\text{Im}[D^{\text{R}}_{\text{ph}}(\omega_{\text{E}})]_{\text{NSC}}/\pi$ is the equilibrium ($F=0$) optical phonon DOS in the NSC case, see Eq.~\eqref{eq:local_Dyson_Ein_ph}. On the other hand, in setup A the phonon coupling strength reads $\Gamma_{\text{ph}} \equiv 2\pi g^{2}\rho_{\text{D}}(\omega_{\text{D}})$, with the acoustic phonon DOS $\rho_{\text{D}}(\omega_{\text{D}})$ given in Eq.~\eqref{eq:ac_ohmic_DOS}.}
    \label{tab:default_pars}
  \end{center}
\end{table}

\section{Results}\label{sec:results}

We study a Mott insulating system \resub{with $U=8t^{\ast}$ (which, as it has been pointed out in Ref.~\cite{mu.we.18}, is already enough to have a well-established gap in the infinite-dimensional case)}, attached to an electron bath \resub{by means of the electronic decay rate} $\Gamma_{\text{e}}$, see Eq.~\eqref{eq:WBL_bathGF}, {\em plus} optical (setup O) or acoustic (setup A) phonons. The phonon coupling strength $\Gamma_{\text{ph}}$ for both optical and acoustic phonons is defined in Tab.~\ref{tab:default_pars}, while the \resub{electronic decay rate} reads $\Gamma_{\text{e}}\equiv 2\pi v^{2}S_{\text{bath}}(0)$ with $S_{\text{bath}}(\omega) = -\text{Im}[g^{\text{R}}_{\text{bath}}(\omega)]/\pi$, see Eq.~\eqref{eq:WBL_bathGF}. In this paper we choose $\Gamma_{\text{e}}/t^{\ast}=\left\{ 0.12, 0.16, 0.20 \right\}$ and set the temperature of the fermionic bath equal to that of the phonon one. If not stated otherwise, the parameters in Tab.~\ref{tab:default_pars} will be used. Setting $\kel{\Pi}_{\text{e-ph}}$ to zero in Eqs~\eqref{eq:local_Dyson_ac_ph} and~\eqref{eq:local_Dyson_Ein_ph} corresponds to the NSC scheme, as opposed to the SC one.

\subsection{Optical phonons}\label{sec:AMEA_Ein_ph}

We first discuss the case corresponding to the setup O in Tab.~\ref{tab:default_pars}, in which the system is coupled to fermionic baths and optical phonons, see Eqs~\eqref{eq:local_Dyson_Ein_ph} and \eqref{eq:non-int_einstein_ph}.

\subsubsection{Current, energy, double occupation}\label{sec:Ein_ph_obs}

\begin{figure}[t]
\includegraphics[width=\linewidth]{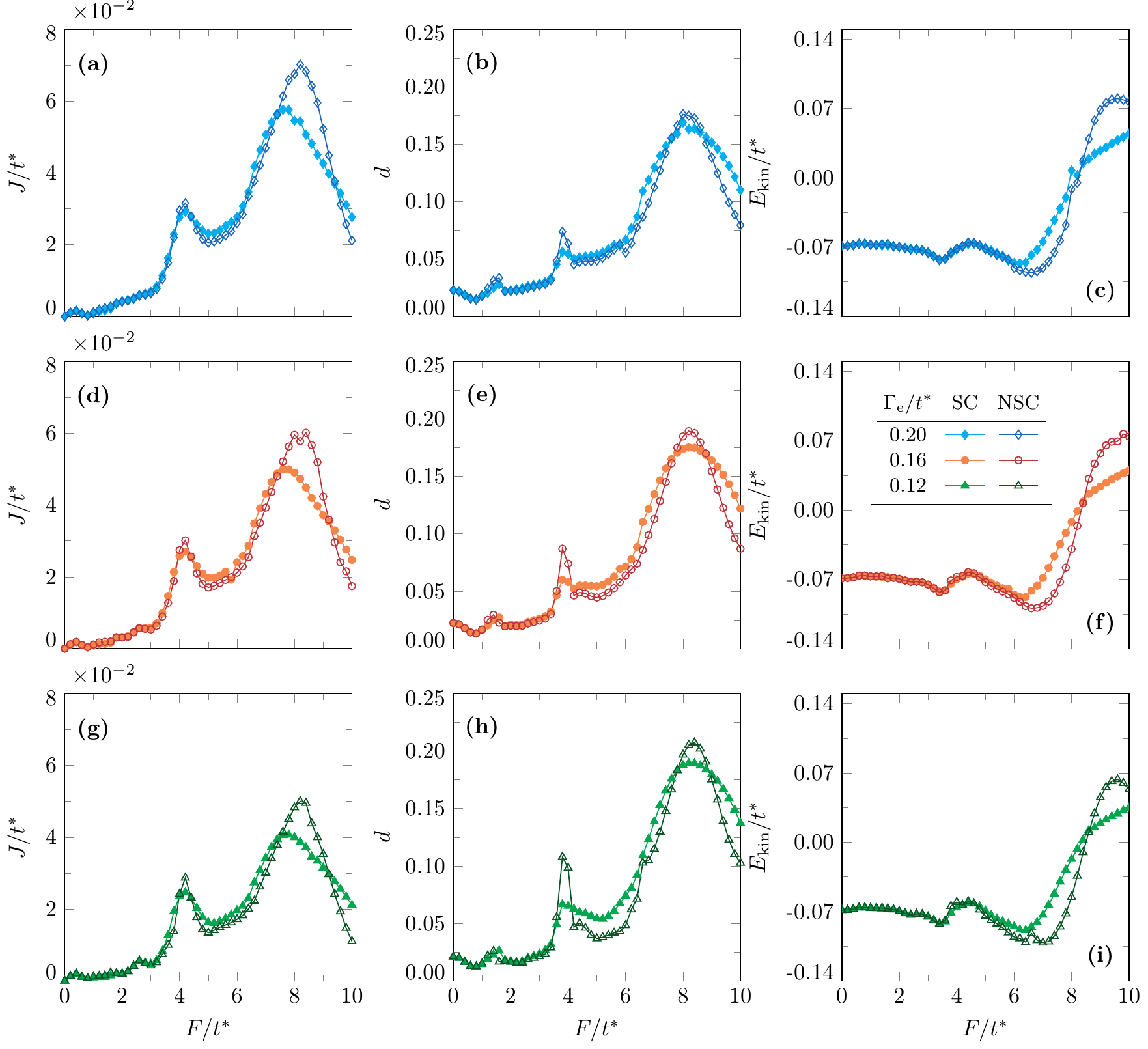}
\caption{\resub{(a) Current $J$, (b) double occupation $d$ and (c) kinetic energy $E_{\text{kin}}$ as function of the applied field $F$ in the SC and NSC schemes for $\Gamma_{\text{e}}=0.2t^{\ast}$. Panels (d) to (f) show the same quantities for $\Gamma_{\text{e}}=0.16t^{\ast}$ while (g) to (i) for $\Gamma_{\text{e}}=0.12t^{\ast}$. Default parameters refer to the optical phonon case (setup O) in Tab.~\ref{tab:default_pars}. (Here $U=8t^{\ast}$.)}}
\label{fig:observables_Hols}
\end{figure}

The current $J$, double occupation per site $d$ and kinetic energy $E_{\text{kin}}$ as function of the applied field $F$ for selected \resub{electronic decay rates} $\Gamma_{\text{e}}$~\footnote{We choose the values of the \resub{electronic decay rate} $\Gamma_{\text{e}}$ such to ensure that a {\em stable} steady-state in the sense discussed in Ref.~\cite{ma.ga.22} is reached.} are shown in Fig.~\ref{fig:observables_Hols}. 

Regardless of the value of $\Gamma_{\text{e}}$, the two resonances at $F\approx U/2=4t^{\ast}$ and $F\approx U=8t^{\ast}$ in $J$ are accompanied by enhancements in $d$~\cite{ma.ga.22,mu.we.18}, as evidenced by Figs~\ref{fig:observables_Hols}(a), (b) and (c) for the former and (d), (e) and (f) for the latter. On the other hand, the kinetic energy $E_{\text{kin}}$ shows a {\em plateau}-like behavior followed by an inflection point at around $F\approx 4t^{\ast}$ and then rises sharply, starting from $F\approx 7t^{\ast}$ as shown in Figs~\ref{fig:observables_Hols}(g), (h) and (i). Note that $E_{\text{kin}}$ keeps growing even at field strengths at which both $d$ and $J$ are already suppressed. This signals that the injected energy no longer increases the mobility of the electrons but rather promotes their incoherent motion~\footnote{For $F>U$ any further increase in the applied field no longer results in a net motion of charge carriers, thus the injected energy only increases the systems temperature.}. These findings are qualitatively robust against the value of $\Gamma_{\text{e}}$: as shown in Ref.~\cite{ma.ga.22}, a larger \resub{electronic decay rate} is more effective in relaxing excited charge carriers to the lower Hubbard band (LHB), which implies a reduction of the double occupancy at $F\approx U$, see Figs~\ref{fig:observables_Hols}(d), (e) and (f). We see that in the region $F<U/2$ the observables are basically identical in the SC and NSC schemes, while for $F\in [U/2,U]$ and $F>U$ both $J$ and $d$ are slightly enhanced by the SC treatment. This is in contrast with the regions around the resonances $F\approx U/2$ and $F\approx U$, in which the values of the current, double occupancy and kinetic energy are reduced in the SC scheme. Finally we note that within the SC treatment the sharp increase of $E_{\text{kin}}$ observed in Fig.~\ref{fig:observables_Hols} gets mitigated, even though its tendency is preserved. This reduction may suggest an energy transfer from electrons to phonons in which the latter {\em absorb} part of the kinetic energy from the former in the form of heat. We will further develop this aspect at the end of Sec.~\ref{sec:Ein_ph_spec} by analyzing the phonon spectra.

\begin{figure}[t]
\includegraphics[width=\linewidth]{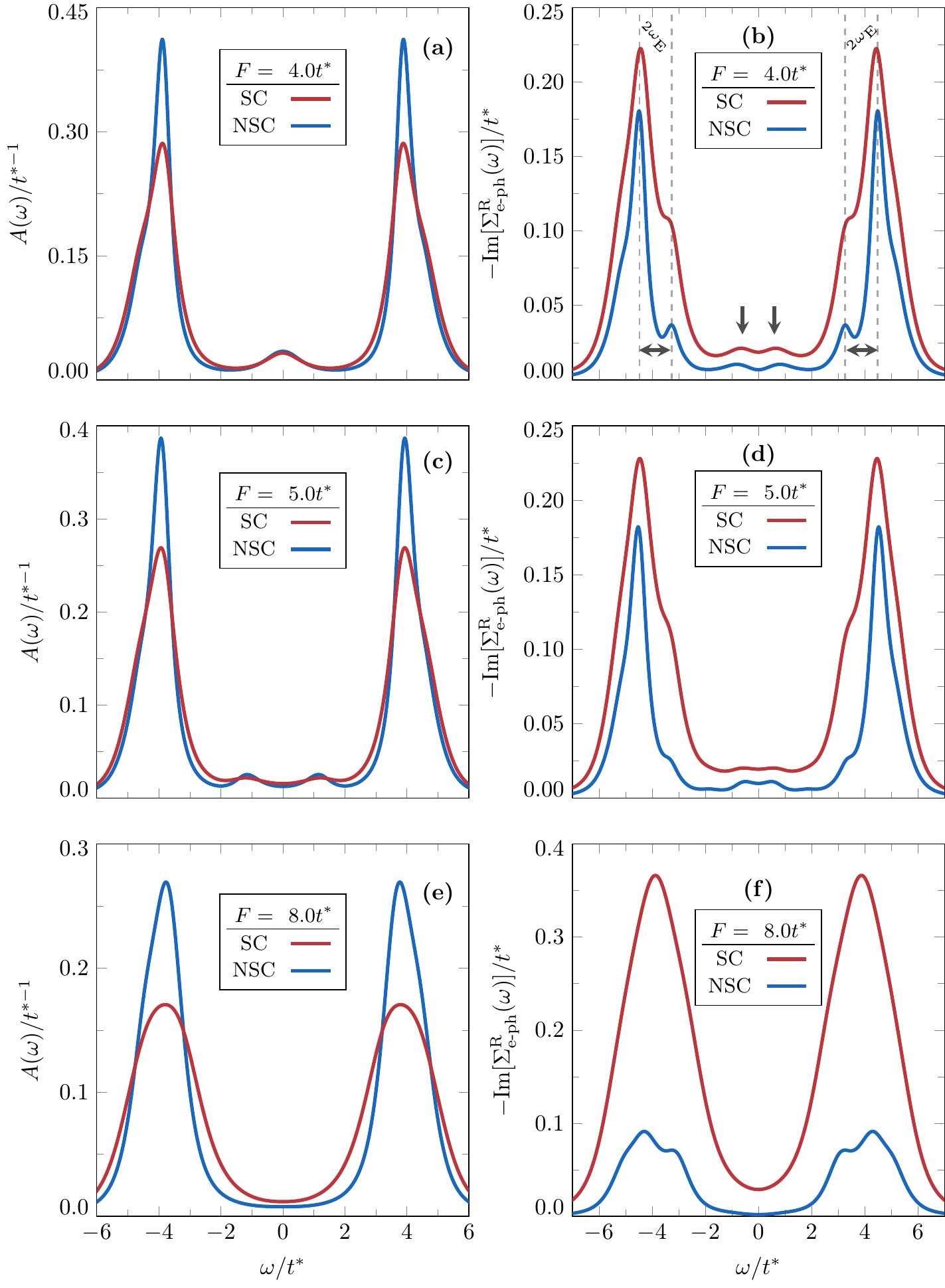}
\caption{(a) Electron spectral function $A(\omega)$ and (b) e-ph SE $-\text{Im}\Sigma^{\text{R}}_{\text{e-ph}}(\omega)$ at $F=4t^{\ast}$. Black vertical arrows in (b) point at the in-gap peaks at $\omega\approx \pm\omega_{\text{E}}$, while the horizontal ones highlight the separation $\delta\approx 2\omega_{\text{E}}$ between the subpeaks in which the main bands are split. Panels (c) and (d) show the same quantities at $F=5t^{\ast}$, while (e) and (f) refer to $F=8t^{\ast}$. \resub{Default parameters refer to the optical phonon case (setup O) in Tab.~\ref{tab:default_pars}. (Here $\Gamma_{\text{e}}=0.2t^{\ast}$ and $U=8t^{\ast}$.)}}
\label{fig:EL_SFs_Hols}
\end{figure}

\subsubsection{Spectral properties}\label{sec:Ein_ph_spec}

By the analysis of the spectral properties of both the electrons and phonons we can explain the differences between the SC and NSC schemes observed in Fig.~\ref{fig:observables_Hols}. In particular we focus on the exemplary cases $F=4t^{\ast}$, $F=5t^{\ast}$ and $F=8t^{\ast}$. 

At $F=4t^{\ast}$, Fig.~\ref{fig:EL_SFs_Hols}(a), the electron spectral function $A(\omega)$ shows in-gap states~\cite{aron.12,mu.we.18,ma.ga.22} around $\omega=0$~\footnote{We recall that the in-gap states are due to filled bands of neighboring sites entering the gap of their adjacent ones under the action of the electric field, thus allowing electron tunnelling from the LHB to the UHB.}. These states are accompanied by the filling of the gap in the e-ph SE, Fig.~\ref{fig:EL_SFs_Hols}(b): we observe subpeaks at $\omega\approx \pm\omega_{\text{E}}$ and the splitting of the Hubbard bands~\footnote{Strictly speaking, one should talk of {\em Hubbard bands} only when referring to the electron spectral function $A(\omega)$. In this paper we improperly use the name Hubbard bands also for the main bands in which the e-ph SE is split into. Depending on the context, it should be clear to which objects the authors are referring to.} by an amount $2\omega_{\text{E}}$. This rich structure in the e-ph SE resembles the results in Ref.~\cite{ha.ar.22u}, in which it is argued that electron relaxation across the band gap is mediated by multiple emissions of phonons of energy $\omega_{\text{E}}$. In the SC scheme, the satellite peaks in the e-ph SE are broadened, resulting in an even more pronounced closing of the gap that increases the number of states available to dissipation in the e-ph channel. At $F=5t^{\ast}$ the in-gap states in the electron spectral function are located at $\omega \approx \pm t^{\ast}$, Fig.~\ref{fig:EL_SFs_Hols}(c): the corresponding peak structure in the e-ph SE in Fig.~\ref{fig:EL_SFs_Hols}(d) now gets smeared out in both the SC and NSC schemes. Notice that the SC treatment still leads to a filling of the gap.

\begin{figure}[b]
\includegraphics[width=\linewidth]{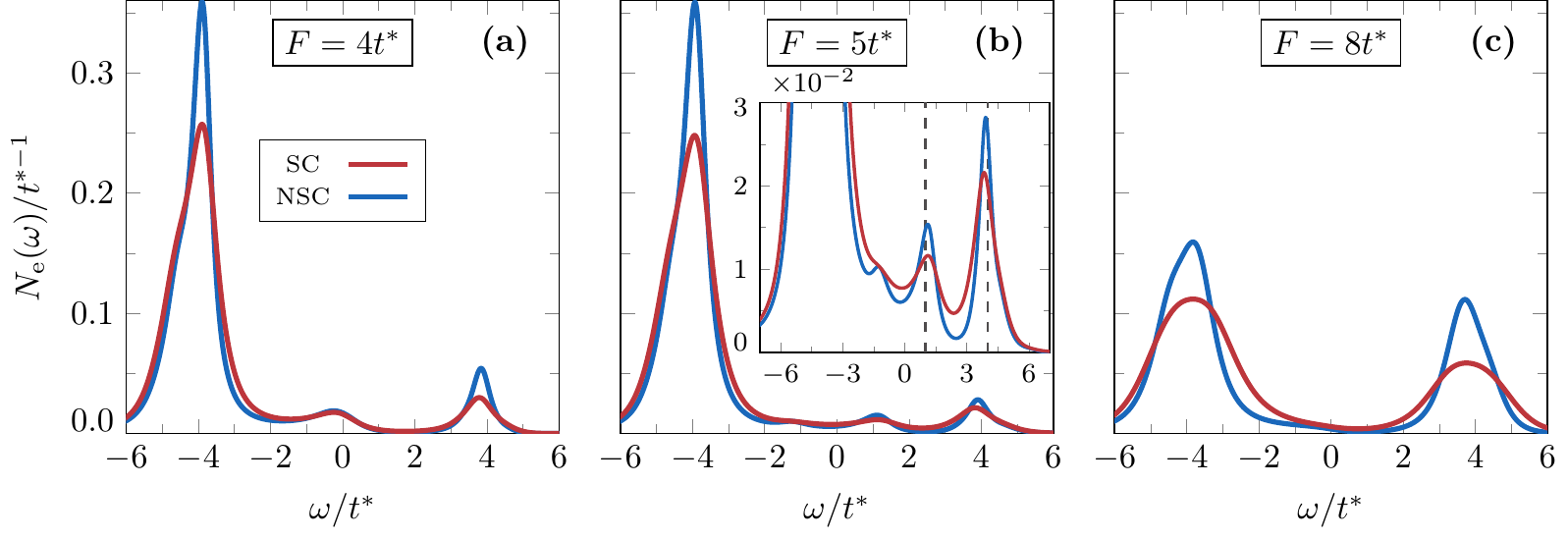}
\caption{The electron spectral occupation function $N_{\text{e}}(\omega)$~\eqref{eq:Filling_func} is shown for field strengths (a) $F=4t^{\ast}$, (b) $F=5t^{\ast}$ and (c) $F=8t^{\ast}$ in the SC and NSC schemes. The inset in (b) magnifies the peaks at $\omega \approx t^{\ast}$ and $\omega \approx 4t^{\ast} = U/2$ (highlighted by vertical dashed black lines). \resub{Default parameters refer to the optical phonon case (setup O) in Tab.~\ref{tab:default_pars}. (Here $\Gamma_{\text{e}}=0.2t^{\ast}$ and $U=8t^{\ast}$.)}}
\label{fig:Hols_EL_occupation_SFs}
\end{figure}

To understand the differences between these two cases, we recall that at $F=U/2$ electrons are promoted to the upper Hubbard band (UHB) through the in-gap states~\cite{ma.ga.22} shown in Fig.~\ref{fig:EL_SFs_Hols}(a). From the electron spectral occupation function $N_{\text{e}}(\omega)$ in Fig.~\ref{fig:Hols_EL_occupation_SFs}(a), we see that in the SC case the occupation of the UHB is reduced. This is due to the increase in the rate of electrons relaxing within the gap via phonon emission --- governed by the in-gap states in Fig.~\ref{fig:EL_SFs_Hols}(b).
The net result is the drop in the current observed in Fig.~\ref{fig:observables_Hols}.
On the other hand, at $F=5t^{\ast}$ electron migration to the UHB requires higher order processes compared to the {\em resonant} transition at $F=U/2$, as evidenced by the in-gap double-peak structure in the spectral function, Fig.~\ref{fig:EL_SFs_Hols}(c). In the SC scheme the occupation of the UHB and of the states around $\omega\approx t^{\ast}$ is reduced as well, see Fig.~\ref{fig:Hols_EL_occupation_SFs}(b) and the inset therein. With the field being off-resonance, the slight enhancement of both $J$ and $d$ noted in Sec.~\ref{sec:Ein_ph_obs} can be attributed to particle flow through the broader in-gap states (Figs~\ref{fig:EL_SFs_Hols}(c), \ref{fig:Hols_EL_occupation_SFs}(b) \resub{and corresponding inset}) induced by the closing of the gap in the e-ph SE shown in Fig.~\ref{fig:EL_SFs_Hols}(d). 

At the resonance $F=U$ the (empty) UHB and the (full) LHB of any pair of neighboring sites match perfectly~\cite{mu.we.18,ma.ga.22}, which explains the absence of in-gap subpeaks in the electron spectral function, see Fig.~\ref{fig:EL_SFs_Hols}(e), and the maximum in the current $J$ observed in Fig.~\ref{fig:observables_Hols}. In the SC scheme, the strong {\em renormalization} of the e-ph SE shown in Fig.~\ref{fig:EL_SFs_Hols}(f) provides the necessary states to relax the electrons within the gap, reducing the occupation of the UHB in favor of the in-gap states, see Fig.~\ref{fig:Hols_EL_occupation_SFs}(c). As in the case of $F=U/2$ this leads to the suppression of the current, double occupation and kinetic energy observed in Fig.~\ref{fig:observables_Hols}.

\begin{figure}[t]
\includegraphics[width=\linewidth]{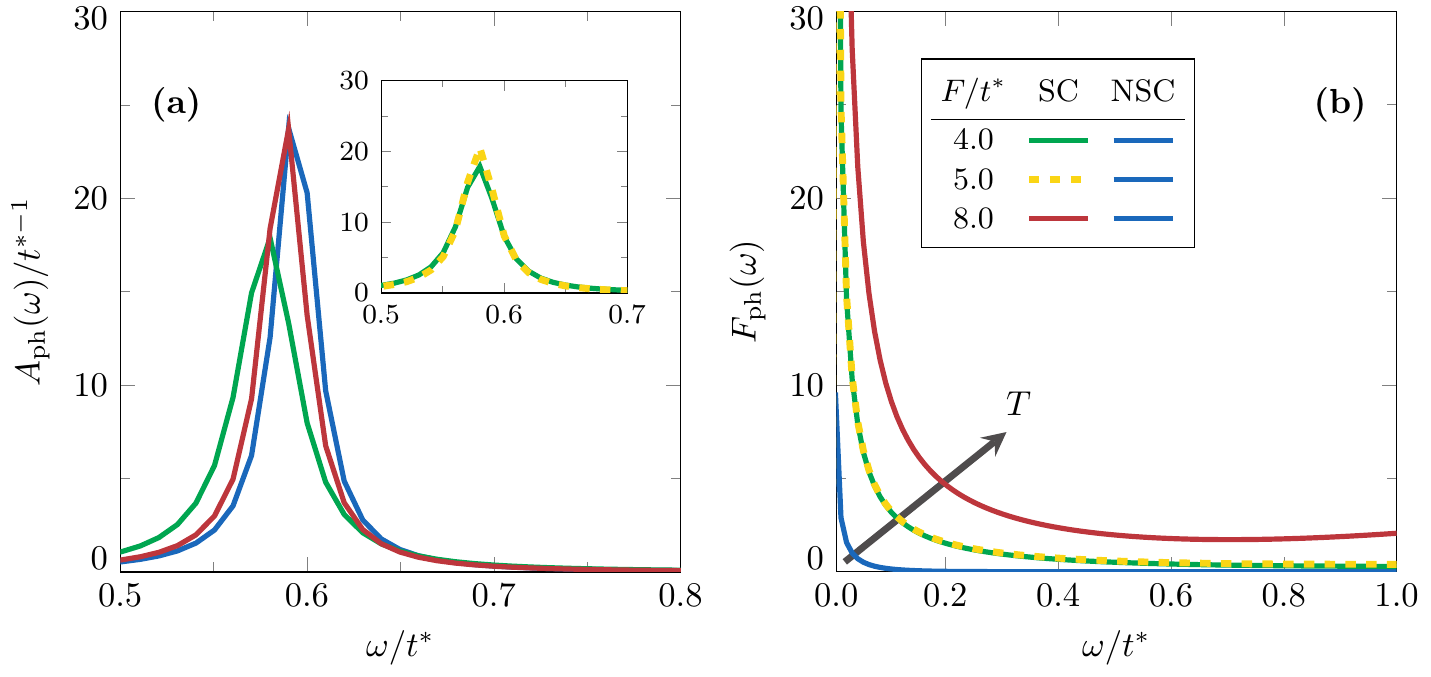}
\caption{(a) Phonon spectral function $A_{\text{ph}}(\omega)$ corresponding to the field strengths shown in Fig.~\ref{fig:EL_SFs_Hols}: the inset compares $F=4t^{\ast}$ and $F=5t^{\ast}$. (c) Nonequilibrium phonon distribution function $F_{\text{ph}}(\omega)$, see Eq.~\eqref{eq:NEFD-dist}, corresponding to (a). The black arrow denotes the direction of increasing temperature. \resub{Default parameters refer to the optical phonon case (setup O) in Tab.~\ref{tab:default_pars}. (Here $\Gamma_{\text{e}}=0.2t^{\ast}$ and $U=8t^{\ast}$.)}}
\label{fig:PH_SFs_Hols}
\end{figure}

In Sec.~\ref{sec:Ein_ph_obs}, we speculated that the increase of $E_{\text{kin}}$ for field strengths at which both $J$ and $d$ are suppressed could be the signature of the injected energy turning into disordered motion of particles, which may eventually lead to an increase in the temperature of the system. While the phonon spectral function in Figs~\ref{fig:PH_SFs_Hols}(a) provides a measure of the {\em renormalization} of the phonon spectrum due to the SC treatment, the phonon nonequilibrium distribution function in Fig.~\ref{fig:PH_SFs_Hols}(b) shows that by increasing the field $F$ phonons experience an increase in temperature. \TMch{It should be noted that, given the non-thermal nature of $F_{\text{ph}}(\omega)$ away from equilibrium, the phonon temperature cannot be inferred by an equally weighted fitting procedure of $F_{\text{ph}}(\omega)$ by means of a Bose-Einstein distribution function --- for more details we refer to the discussion at the end of Sec.~\ref{sec:temp_dep_ac_ph}.} However, given that $F_{\text{ph}}(\omega)$ in Fig.~\ref{fig:PH_SFs_Hols}(b) departs from the equilibrium one as the applied field $F$ grows larger, we can conclude that in the SC scheme the phonon temperature does increase.

\subsubsection{Role of the Hubbard $U$}\label{sec:Hubbard_U_dep}

\begin{figure}[b]
\includegraphics[width=\linewidth]{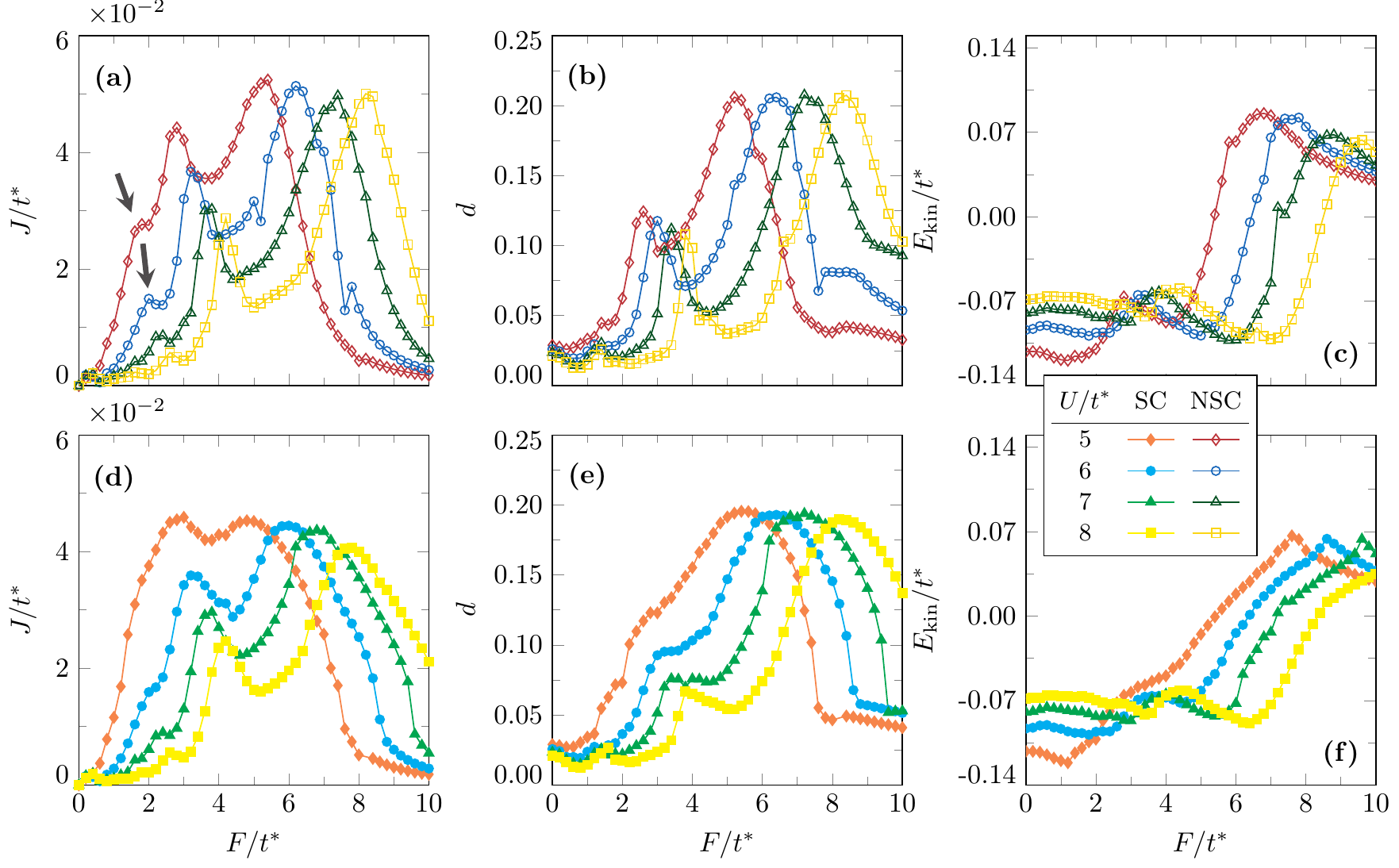}
\caption{(a) Current $J$, (b) double occupation $d$ and (c) kinetic energy $E_{\text{kin}}$ as function of the applied field $F$ for selected values of the Hubbard $U$ in the NSC scheme: black arrows in (a) point at the position of the resonance $F\approx U/3$ which becomes fainter starting at $U=6t^{\ast}$ and turns into an even weaker shoulder at $U=5t^{\ast}$ in the SC scheme, see panel (d). Panels (d), (e) and (f) show the same quantities for the SC case. \resub{Default parameters refer to the optical phonon case (setup O) in Tab.~\ref{tab:default_pars}. (Here $\Gamma_{\text{e}}=0.12t^{\ast}$.)}}
\label{fig:Obs_SC_NSC_Us}
\end{figure}

In the previous section we argued that SC phonons reduce the band gap by relaxing excited carriers from the UHB into it, see Fig.~\ref{fig:EL_SFs_Hols}. Also, it is known that the gap in a single-band Hubbard model at equilibrium ($F=0$) increases as the interaction $U$ grows larger. In this section we want to investigate the effect of SC phonons on a system which exhibits a less pronounced band gap, corresponding to a weaker insulating phase. To this end, we discuss the effects of SC phonons for selected values of the Hubbard interaction $U$. 

As shown in Fig.~\ref{fig:Obs_SC_NSC_Us}, within the NSC scheme we still observe two main resonances at $F\approx U/2$ and $F\approx U$ for both $J$ and $d$, see panels (a) and (b)~\footnote{It should be noted that this double-peak structure in the current and double occupation vanishes as soon as $U$ is too small for the system to develop a band gap. As a matter of fact, below $U=5t^{\ast}$ and with $F=0$ the system does not exhibit a clear gap, thus losing its insulating properties.}. The small resonance at $F\approx U/3$~\cite{mu.we.18,ma.ga.22} can be noticed as well, whereas in the SC scheme the latter gets fainter by decreasing $U$ until it becomes a shoulder to the resonance at $F\approx U/2$ for $U=5t^{\ast}$, see Fig.~\ref{fig:Obs_SC_NSC_Us}(d).

The qualitative difference between the peak (in both $J$ and $d$) at $F\approx U/2$ and $F\approx U$ lies in the fact the the former is reduced by increasing $U$ while the latter approximately preserves its height independently of the value of $U$, see Figs~\ref{fig:Obs_SC_NSC_Us}(a), (b), (d) and (e). This can be explained from the fact that electron transitions from LHB to UHB via in-gap states are suppressed by a larger $U$ if the field strength is off-resonance (as in the case of $F\approx U/2$) while at $F\approx U$ the field allows direct transitions from filled to empty bands, regardless of the value of $U$. Notably, within the NSC scheme the height of the peak in $J$ at $F\approx U$ stays the same, while in the SC treatment the latter changes slightly by varying $U$.

Also, within the SC scheme the peak currents at $F\approx U$ are reduced with respect to the NSC treatment, while $J$ is a bit enhanced away from resonance(s), see for instance Figs~\ref{fig:Obs_SC_NSC_Us}(a) and (d). The double occupation in panels (b) and (e), as well as the kinetic energy, shown in panels (c) and (f), are in agreement with the overall broadening of the $J$-$F$ characteristics, confirming the effects of phonons renormalization discussed in Sec.~\ref{sec:Ein_ph_spec}.
We also notice that the {\em shoulder}-like features occurring in the kinetic energy \resub{near the current conducting state} in the NSC case turn into {\em cusps} in the SC treatment, see Figs~\ref{fig:Obs_SC_NSC_Us} (c) and (f). Another interesting feature is that for small field strengths (i.e. $F\leq 3t^{\ast}$) a larger U translates into a higher kinetic energy while at large fields ($F$ in between the two main resonances) a larger U suppresses the kinetic energy.

\begin{figure}[t]
\includegraphics[width=\linewidth]{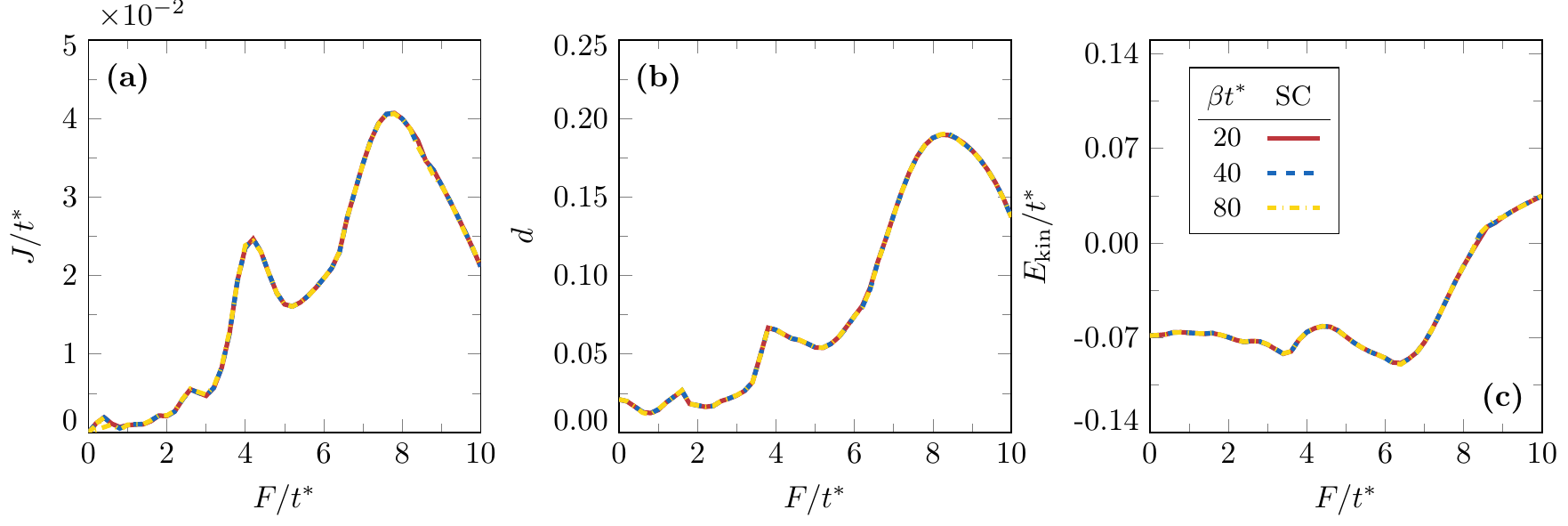}
\caption{(a) Current $J$, (b) double occupation $d$ and (c) kinetic energy $E_{\text{kin}}$ as function of the applied field $F$ for selected values of the inverse temperature $\beta$ within the SC scheme. \resub{Default parameters refer to the optical phonon case (setup O) in Tab.~\ref{tab:default_pars}. (Here $\Gamma_{\text{e}}=0.12t^{\ast}$ and $U=8t^{\ast}$.)}}
\label{fig:Obs_SC_betas}
\end{figure}

Finally a remark on the role of temperature in our simulations. The default inverse temperature has been chosen as $\beta t^{\ast}=20$. As shown in Figs~\ref{fig:Obs_SC_betas}(a), (b) and (c), the current, double occupation and kinetic energy are essentially not affected by lowering the temperature of the electron and phonon bath. This is due the fact that the characteristic frequency of optical phonons $\omega_{\text{E}}$ is typically larger than the temperature of the system.

\subsection{Acoustic phonons}\label{sec:AMEA_ac_ph}

In this section we discuss the influence of acoustic phonons on the electronic properties of the lattice. With a dispersion relation of the form~\eqref{eq:ac_ohmic_DOS}, we expect the cutoff frequency $\omega_{\text{D}}$ to determine the relaxation pathways contributing to heat dissipation. In particular, a smaller $\omega_{\text{D}}$, corresponding to long-wavelength vibrations~\cite{ga.ma.22}, should be more effective in carrying away the heat for longer distances.

In this setup the phonon coupling strength is defined as $\Gamma_{\text{ph}}=2\pi g^{2} \rho_{\text{D}}(\omega_{\text{D}})$, see Sec.~\ref{sec:ac_ph_formalism} and especially Eq.~\eqref{eq:ac_ohmic_DOS}. The default parameters can be found in setup A in Tab.~\ref{tab:default_pars}.

\begin{figure}[t]
\includegraphics[width=\linewidth]{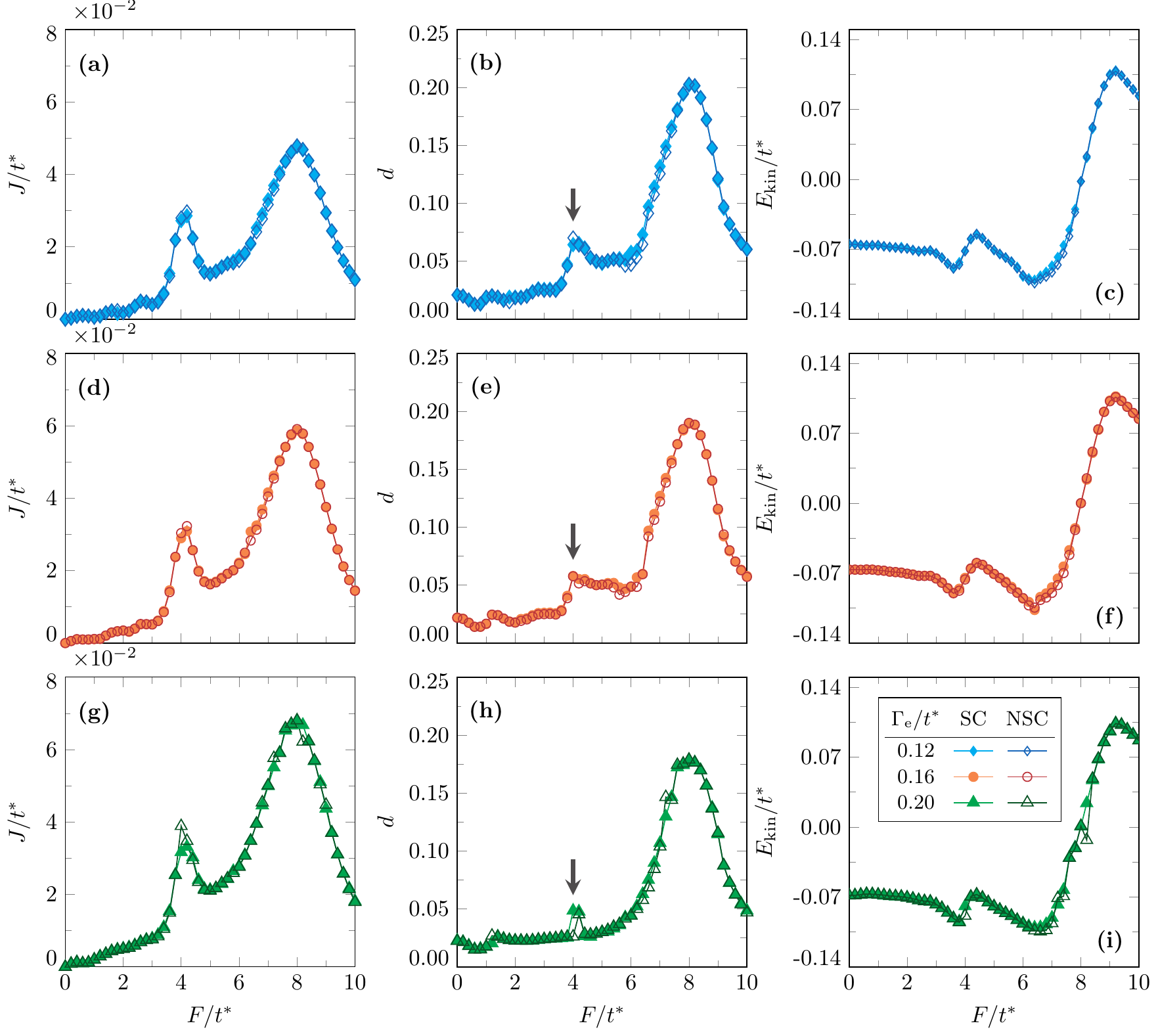}
\caption{(a) Current $J$, (b) double occupation $d$ and (c) kinetic energy $E_{\text{kin}}$ as function of the applied field $F$ for $\Gamma_{\text{e}}=0.12t^{\ast}$. Panels (d) to (f) show the same quantities for $\Gamma_{\text{e}}=0.16t^{\ast}$, while the results for $\Gamma_{\text{e}}=0.2t^{\ast}$ are displayed in panels (g) to (i). Black arrows highlight the suppression of the resonance at $F\approx U/2$ in $d$ as function of the increasing $\Gamma_{\text{e}}$. \resub{Default parameters refer to the acoustic phonon case (setup A) in Tab.~\ref{tab:default_pars}. (Here $U=8t^{\ast}$.)}}
\label{fig:ac_ph_obs_Gammaes}
\end{figure}

\subsubsection{Current, energy and double occupation}\label{sec:coup_ferm_bath_ac_ph}

\paragraph{Role of the \resub{electronic decay rate} $\Gamma_{\text{e}}$.}\label{sec:role_el_bath}

We start from the analysis of $J$, $d$ and $E_{\text{kin}}$: in Fig.~\ref{fig:ac_ph_obs_Gammaes} these quantities are shown as function of the applied field $F$ for selected values of the \resub{electronic decay rate} $\Gamma_{\text{e}}$. Let alone the small suppression of $J$ at around $F\approx U/2$ in the SC case, which is visible in Figs~\ref{fig:ac_ph_obs_Gammaes}(a), (d) and especially (g), there are no remarkable differences between the SC and NSC schemes as the observables almost lie on top of each other, see panels (b), (e) and (h) for $d$ and (c), (f) and (i) for $E_{\text{kin}}$. This is in contrast with the case of optical phonons discussed in Sec.~\ref{sec:AMEA_Ein_ph}, where the effects of the SC treatment were plainly visible.

For later purposes, we just want to stress the suppression of the resonant peak in the double occupation at $F\approx U/2$, see Figs~\ref{fig:ac_ph_obs_Gammaes}(b), (e) and (h), within both the SC and NSC schemes as $\Gamma_{\text{e}}$ is increased.

\paragraph{Role of the phonon cutoff frequency $\omega_{\text{D}}$.}\label{sec:obs_ac_ph}

\begin{figure}[b]
\includegraphics[width=\linewidth]{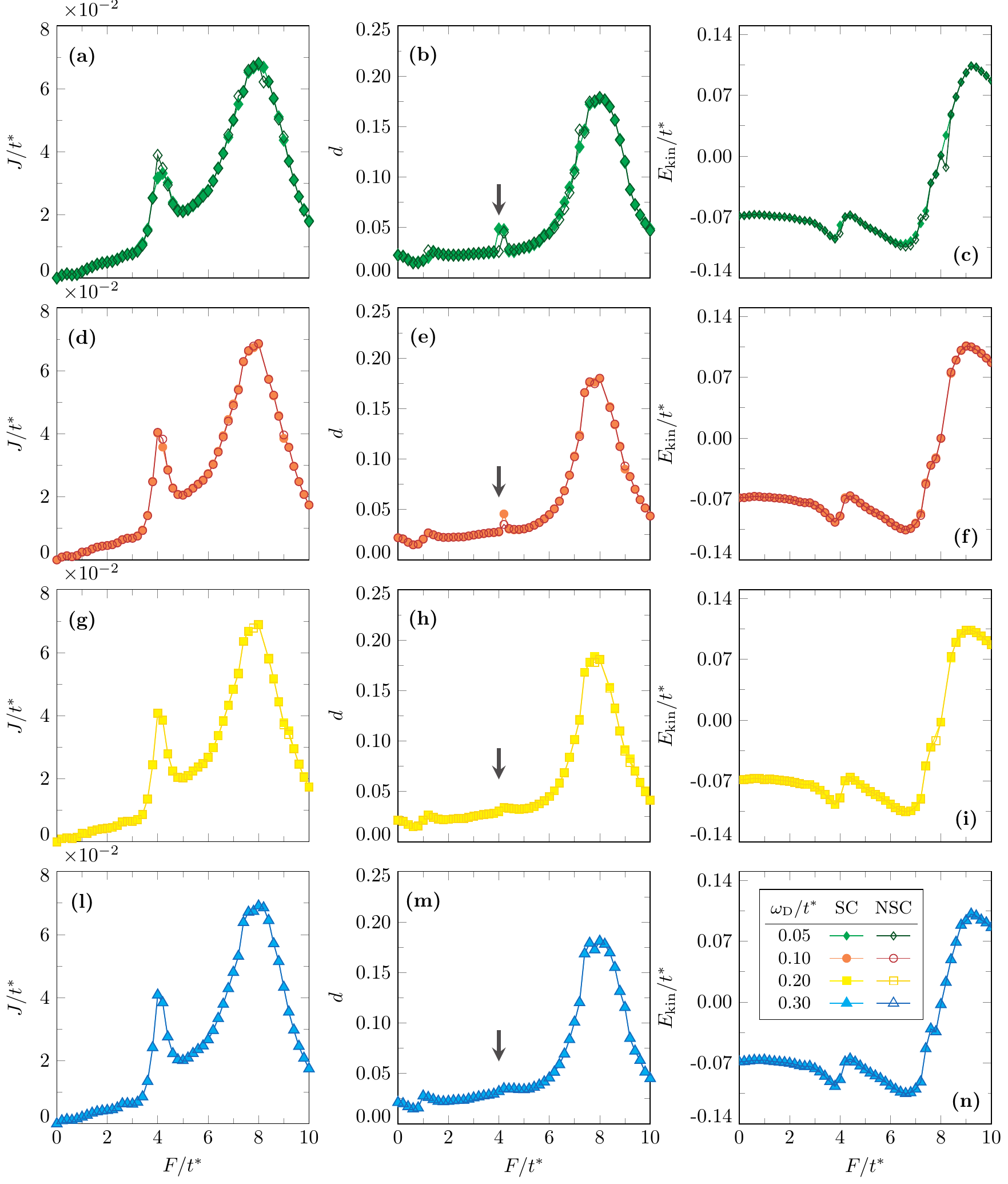}
\caption{(a) Current $J$, (b) double occupation $d$ and (c) kinetic energy $E_{\text{kin}}$ as function of the applied field $F$ for $\omega_{\text{D}}=0.05t^{\ast}$ within the NSC and SC schemes. Panels (d) to (f) show the same quantities for $\omega_{\text{D}}=0.1t^{\ast}$, (g) to (i) for $\omega_{\text{D}}=0.2t^{\ast}$, while the results for $\omega_{\text{D}}=0.3t^{\ast}$ are displayed in panels (l) to (n). The black arrows highlight the suppression of the resonance at $F\approx U/2$ in $d$ as $\omega_{\text{D}}$ is increased. \resub{Default parameters refer to the acoustic phonon case (setup A) in Tab.~\ref{tab:default_pars}. (Here $\Gamma_{\text{e}}=0.2t^{\ast}$ and $U=8t^{\ast}$.)}}
\label{fig:Obs_ac_ph}
\end{figure}

In Fig.~\ref{fig:Obs_ac_ph} we show $J$, $d$ and $E_{\text{kin}}$ as function of the applied field for $\Gamma_{\text{e}}=0.2t^{\ast}$ and selected values of $\omega_{\text{D}}$. As noted in Sec.~\ref{sec:role_el_bath}, at $\omega_{\text{D}}=0.05t^{\ast}$ the current $J$ is slightly suppressed at $F\approx U/2$ within the SC scheme, see Fig.~\ref{fig:Obs_ac_ph}(a). However, the $J$-$F$ curve does not show appreciable changes for all the other $\omega_{\text{D}}$'s used in this paper, see Figs~\ref{fig:Obs_ac_ph}(d), (g) and (l). On the other hand, the resonance at $F\approx U/2$ in the double occupation $d$ is suppressed by increasing the phonon cutoff frequency $\omega_{\text{D}}$ within both the SC and NSC schemes as it is clear from Figs~\ref{fig:Obs_ac_ph}(b), (e), (h) and (m). Notably, such an effect occurs already for increasing coupling $\Gamma_{\text{e}}$, again see the discussion in~\ref{sec:role_el_bath}. Finally, the kinetic energy is not affected by changing the soft cutoff frequency $\omega_{\text{D}}$ or by the SC scheme, as one can see by direct inspection of Figs~\ref{fig:Obs_ac_ph}(c), (f), (i) and (n).

\subsubsection{Spectral properties}\label{sec:spectral_prop_ac_ph}

To explain the findings discussed in Sec.~\ref{sec:coup_ferm_bath_ac_ph}, we study the spectral properties of both the electrons and phonons. 

\begin{figure}[t]
\includegraphics[width=\linewidth]{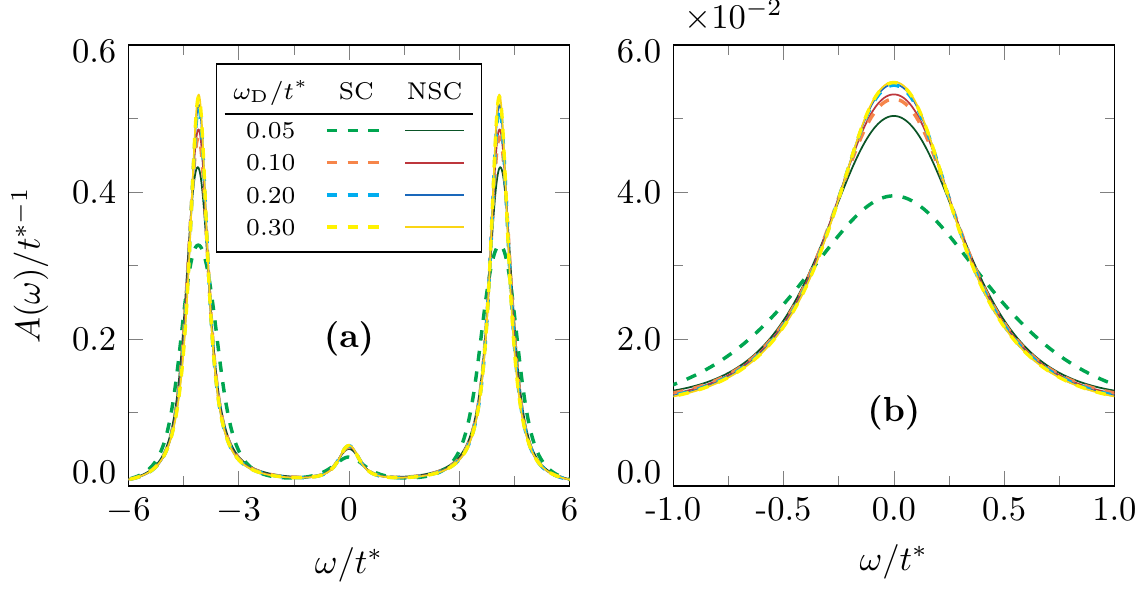}
\caption{(a) Electron spectral function $A(\omega)$ at $F=4t^{\ast}$ for selected values of the phonon cutoff frequency $\omega_{\text{D}}$ within the SC and NSC schemes. Panel (b) magnifies the {\em quasi-particle} peak at $\omega\approx 0$. \resub{Default parameters refer to the acoustic phonon case (setup A) in Tab.~\ref{tab:default_pars}. (Here $\Gamma_{\text{e}}=0.2t^{\ast}$ and $U=8t^{\ast}$.)}}
\label{fig:EL_SFs_ac_ph}
\end{figure}

In Fig.~\ref{fig:EL_SFs_ac_ph}(a) the electron spectral function $A(\omega)$ is shown at $F= U/2$ for several values of the phonon cutoff frequency $\omega_{\text{D}}$, from which we see that the SC scheme does not alter the overall electron spectral properties. On the other hand, looking at the low-energy region $\omega\approx 0$ in Fig.~\ref{fig:EL_SFs_ac_ph}(b) we notice appreciable differences in the {\em quasi-particle} peak therein especially at $\omega_{\text{D}}=0.05t^{\ast}$. As a matter of fact, for $\omega_{\text{D}}/t^{\ast}=\left\{ 0.2,0.3 \right\}$ the quasi-particle peak at $\omega \approx 0$ is essentially unaltered by the SC treatment, while a slight suppression (within the SC scheme) can be detected starting from $\omega_{\text{D}}=0.1t^{\ast}$.

As already pointed out in this paper, when the applied field is far from the main resonance ($F\approx U$) the in-gap spectral weight contributes to excite particles from the LHB to the UHB. In this framework, the suppression of the {\em quasi-particle} peak at $\omega \approx 0$ that occurs in the SC scheme at $\omega_{\text{D}}=0.05t^{\ast}$ signals that fewer states are available within the gap, with the consequent reduction of the current $J$ with respect to the NSC case that shown in Fig.~\ref{fig:Obs_ac_ph}(a). This is further proved from the fact that the current $J$ is not affected by the SC treatment in all the other cases $\omega_{\text{D}}/t^{\ast}=\left\{ 0.1,0.2,0.3 \right\}$, see Figs~\ref{fig:Obs_ac_ph}(d), (g) and (l), for which there is no reduction in the quasi-particle peak at $\omega \approx 0$ as shown in Fig.~\ref{fig:EL_SFs_ac_ph}(b).

\begin{figure}[t]
\includegraphics[width=\linewidth]{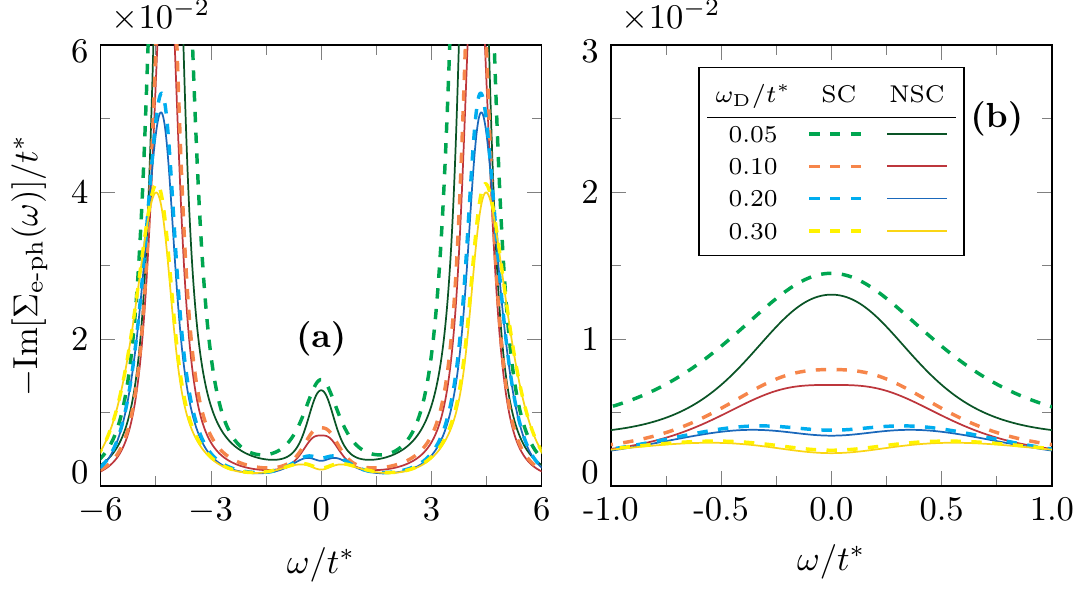}
\caption{(a) Imaginary part of the electron-phonon self-energy $\text{Im}[\Sigma_{\text{e-ph}}(\omega)]$ at $F=4t^{\ast}$ for selected values of the phonon cutoff frequency $\omega_{\text{D}}$ within the SC and NSC schemes. Panel (b) magnifies the {\em quasi-particle} peak at $\omega\approx 0$. \resub{Default parameters refer to the acoustic phonon case (setup A) in Tab.~\ref{tab:default_pars}. (Here $\Gamma_{\text{e}}=0.2t^{\ast}$ and $U=8t^{\ast}$.)}}
\label{fig:eph_SEs_ac_ph}
\end{figure}
\begin{figure}[b]
\includegraphics[width=\linewidth]{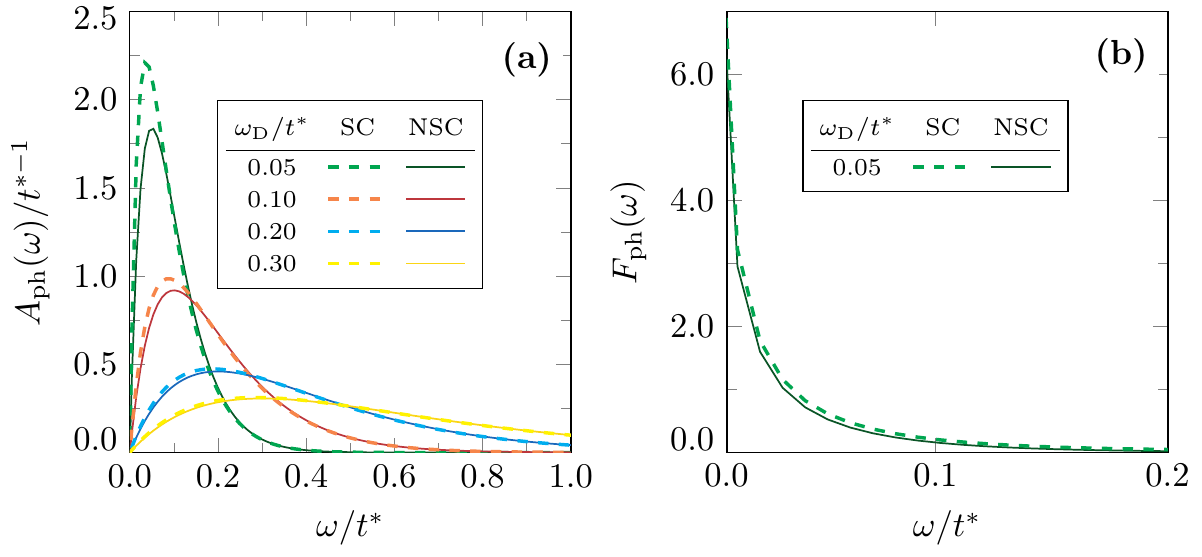}
\caption{(a) Phonon spectral function $A_{\text{ph}}(\omega)$ at $F=4t^{\ast}$ for selected values of $\omega_{\text{D}}$. Panel (b) shows the nonequilibrium phonon distribution function $F_{\text{ph}}(\omega)$ at $F=4t^{\ast}$ and $\omega_{\text{D}}=0.05t^{\ast}$, where the difference between SC and NSC schemes can be appreciated. \resub{Default parameters refer to the acoustic phonon case (setup A) in Tab.~\ref{tab:default_pars}. (Here $\Gamma_{\text{e}}=0.2t^{\ast}$ and $U=8t^{\ast}$.)}}
\label{fig:ph_SFs_ac_ph}
\end{figure}
The e-ph SE at $F=4t^{\ast}=U/2$ is shown in Fig.~\ref{fig:eph_SEs_ac_ph}(a): we observe that the smaller $\omega_{\text{D}}$ the higher the in-gap peak in $\text{Im}[\Sigma^{\text{R}}_{\text{e-ph}}(\omega)]$, see also the magnification of the low-energy region $\omega \approx 0$, Fig.~\ref{fig:eph_SEs_ac_ph}(b), especially for $\omega_{\text{D}}=0.05t^{\ast}$. Also, it should be noted that the height of the peak at $\omega \approx 0$ is always larger in the SC than in the NSC treatment, as opposed to what happens in the electron spectral function, Fig.~\ref{fig:EL_SFs_ac_ph}(b). 

Also, while with optical phonons the increase in the in-gap spectral weight in $\text{Im}[\Sigma^{\text{R}}_{\text{e-ph}}(\omega)]$ is accompanied by an increase in $A(\omega)$, see Fig.~\ref{fig:EL_SFs_Hols} in Sec.~\ref{sec:Ein_ph_spec}, when considering acoustic phonons an increase in the height of the in-gap states in the e-ph SE is characterized by fewer states available in the electron spectral function, see Figs~\ref{fig:EL_SFs_ac_ph}(b) and~\ref{fig:eph_SEs_ac_ph}(b).

It is worth stressing that the height of the in-gap peak in $\text{Im}[\Sigma^{\text{R}}_{\text{e-ph}}(\omega)]$ is suppressed and the latter is split by increasing the cutoff phonon frequency $\omega_{\text{D}}$, see again Fig.~\ref{fig:eph_SEs_ac_ph}(b).

In Fig.~\ref{fig:ph_SFs_ac_ph}(a) we compare the phonon spectral function $A_{\text{ph}}(\omega)$ at $F=4t^{\ast}$ for different values of the phonon cutoff frequency $\omega_{\text{D}}$. The SC treatment shifts the phonon cutoff frequency towards smaller values and increases the height of the phonon spectral function the more the smaller $\omega_{\text{D}}$. In fact, the phonon spectral functions in the SC and NSC schemes tend to coincide as $\omega_{\text{D}}$ is increased. As already pointed out, one can argue that by decreasing $\omega_{\text{D}}$ acoustic phonons should be more effective in carrying away the current-induced heat from the lattice due to their long-wavelength character. In this framework the suppression of $J$ at $F\approx U/2$ for $\omega_{\text{D}}=0.05t^{\ast}$, see Fig.~\ref{fig:Obs_ac_ph}(a), can be explained as the result of the increased spectral weight in both $A_{\text{ph}}(\omega)$ and $\text{Im}[\Sigma^{\text{R}}_{\text{e-ph}}(\omega)]$. Also, the phonon temperature in the SC and NSC schemes is almost the same for this case, as it can be inferred from the nonequilibrium distribution function $F_{\text{ph}}(\omega)$ in Fig.~\ref{fig:ph_SFs_ac_ph}(b). We remark that for the other values of $\omega_{\text{D}}$ used in this paper there are no appreciable changes in the nonequilibrium phonon distribution function $F_{\text{ph}}(\omega)$ between the SC and NSC schemes.

\subsubsection{Temperature dependence}\label{sec:temp_dep_ac_ph}

In this section we discuss the dependence of the above results on the temperature of the electron and phonon baths. 

\begin{figure}[h]
\includegraphics[width=\linewidth]{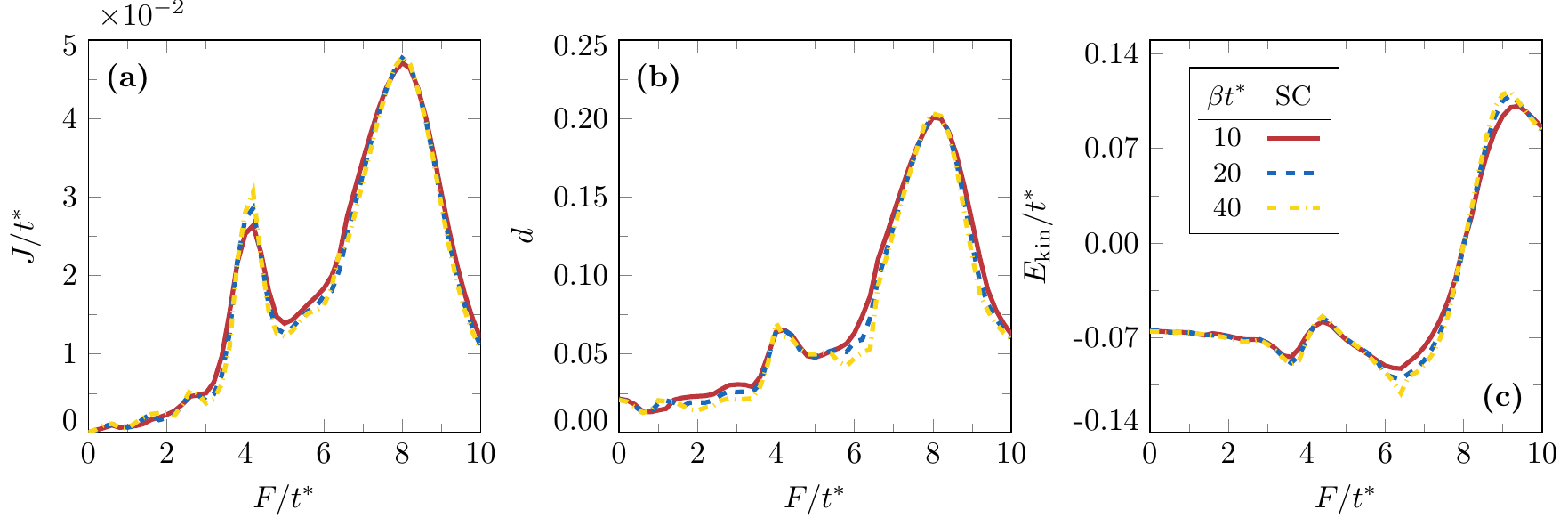}
\caption{(a) Current $J$, (b) double occupation $d$ and (c) kinetic energy $E_{\text{kin}}$ as function of the applied filed $F$ for selected values of the inverse temperature $\beta$ within the SC scheme. Lowering the temperature leads to a detachment of the current curves especially at the resonance $F\approx U/2$. \resub{Default parameters refer to the acoustic phonon case (setup A) in Tab.~\ref{tab:default_pars}. (Here $\Gamma_{\text{e}}=0.12t^{\ast}$ and $U=8t^{\ast}$.)}}
\label{fig:ac_ph_obs_temp_dep}
\end{figure}

In Fig.~\ref{fig:ac_ph_obs_temp_dep} we show the current $J$, double occupation $d$ and kinetic energy $E_{\text{kin}}$ as function of the applied field $F$ for selected values of the inverse temperature $\beta$ within the SC scheme. We see that in contrast to the case of optical phonons, see Fig.~\ref{fig:Obs_SC_betas}, here the curves differ appreciably. In particular, at $F\approx U/2$ the current $J$ is suppressed by increasing the temperature, see Fig.~\ref{fig:ac_ph_obs_temp_dep}(a), while at the same field strength both $d$ and $E_{\text{kin}}$ are essentially not altered, see Figs~\ref{fig:ac_ph_obs_temp_dep}(b) and (c).

\begin{figure}[t]
\includegraphics[width=\linewidth]{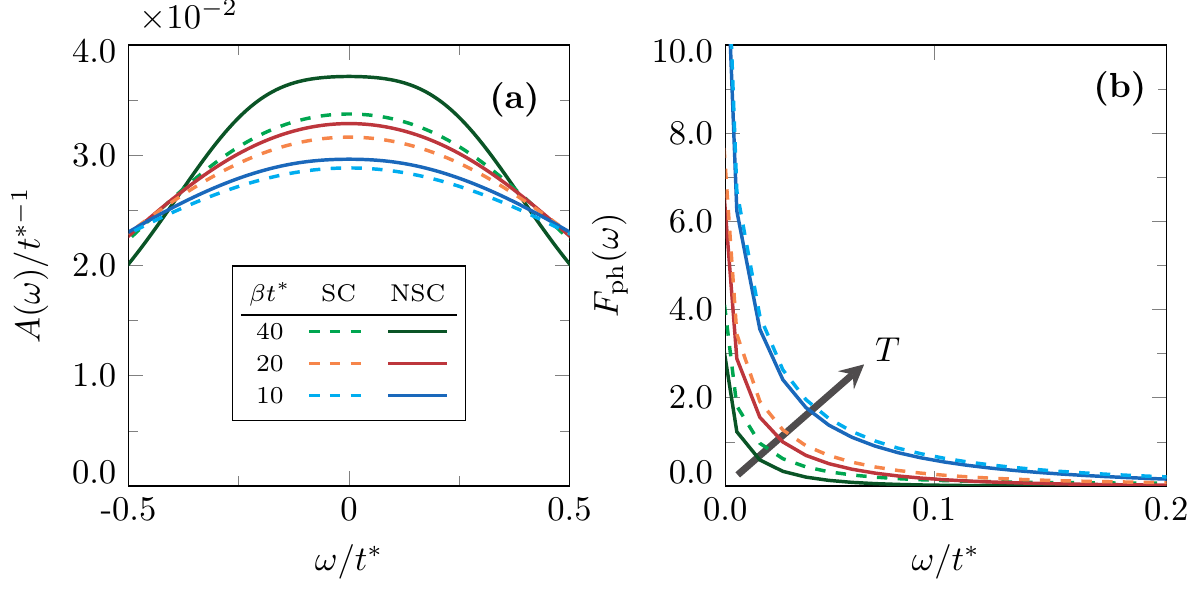}
\caption{(a) Quasi-particle peak at $\omega\approx 0$ in the electron spectral function $A(\omega)$ at field strength $F=4.2t^{\ast}$ for selected values of the inverse temperature $\beta$. (b) Nonequilibrium phonon distribution function $F_{\text{ph}}(\omega)$ corresponding to the same situation as in panel (a). The black arrow denotes the direction of increasing temperature. \resub{Default parameters refer to the acoustic phonon case (setup A) in Tab.~\ref{tab:default_pars}. (Here $\Gamma_{\text{e}}=0.12t^{\ast}$ and $U=8t^{\ast}$.)}}
\label{fig:ac_ph_SF+DF_temp}
\end{figure}

Finally, in Fig.~\ref{fig:ac_ph_SF+DF_temp}, we show the electron spectral function $A(\omega)$ and the nonequilibrium phonon distribution function $F_{\text{ph}}(\omega)$ at $F\approx U/2$ for several values of the inverse temperature $\beta$. One can observe the suppression of the quasi-particle peak at $\omega \approx 0$ as the temperature is increased in both the SC and NSC cases, see panel (a). Also, the SC scheme always reduces the in-gap spectral weight of the electrons with respect to the NSC scheme, see again Fig.~\ref{fig:ac_ph_SF+DF_temp}(a). It should be noted that the largest difference in the height of the quasi-particle peak between the SC and NSC cases occurs at $\beta t^{\ast}= 40$.

Figure~\ref{fig:ac_ph_SF+DF_temp}(b) shows that the phonons experience temperature increase within the SC scheme; in relative terms, the temperature change between SC and NSC treatments is the largest at $\beta t^{\ast}=40$ --- notice the difference in the distribution function $F_{\text{ph}}(\omega)$ between the two cases in Fig.~\ref{fig:ac_ph_SF+DF_temp}(b).

\resub{Finally, we would like to characterize the temperature increase experienced by both acoustic and optical phonons at self-consistency by means of the notion of {\em effective temperature} as it has been introduced in~\cite{ha.ar.22u,di.ha.22u}. Therein they propose the following expression for the phonon effective temperature}
\begin{equation}\label{eq:proposed_eff_temp}
 T^{2}_{\text{ph}} = \frac{6}{\pi^{2}} \int_{0}^{\infty} \dd\omega \ \omega F_{\text{ph}}(\omega).
\end{equation}

\resub{However, plugging our expression for the nonequilibrium phonon distribution function~\eqref{eq:NEBD-dist} into Eq.~\eqref{eq:proposed_eff_temp} yields unstable results due to the ratio $\text{Im}D^{\text{K}}_{\text{ph}}/2\text{Im}D^{\text{R}}_{\text{ph}}$, which has large fluctuations in the frequency regions where the phonon spectral function $A_{\text{ph}}$ is small.}

\TMch{On the other hand, as already discussed at the end of Sec.~\ref{sec:Ein_ph_spec}, given the non-thermal nature of the steady-state reached by the system, fitting $F_{\text{ph}}$ with an equally weighted Bose distribution $f_{\text{B}}(\omega,\beta^{\text{eff}}_{\text{ph}})=1/\left[\ee^{\beta^{\text{eff}}_{\text{ph}}\omega} - 1\right]$ is not a viable option.}

\TMch{For this reason, in our case it is more appropriate to introduce the following {\em cost function}}
\begin{equation}\label{eq:ph_cost_function}
 C \equiv \int_{0}^{\infty} \dd\omega \ \omega A_{\text{ph}}(\omega) |F_{\text{ph}}(\omega)-f_{\text{B}}(\omega,\beta^{\text{eff}}_{\text{ph}})|^{2},
\end{equation}
\TMch{
and determine the phonon effective inverse temperature $\beta^{\text{ph}}_{\text{eff}}=1/T^{\text{eff}}_{\text{ph}}$ from its minimum.
As one can see by direct inspection, Eq.~\eqref{eq:ph_cost_function} introduces a weighted fitting procedure by means of the phonon spectral function $A_{\text{ph}}$, penalizing the low-frequency region.}

\begin{figure}[t]
 \includegraphics[width=\linewidth]{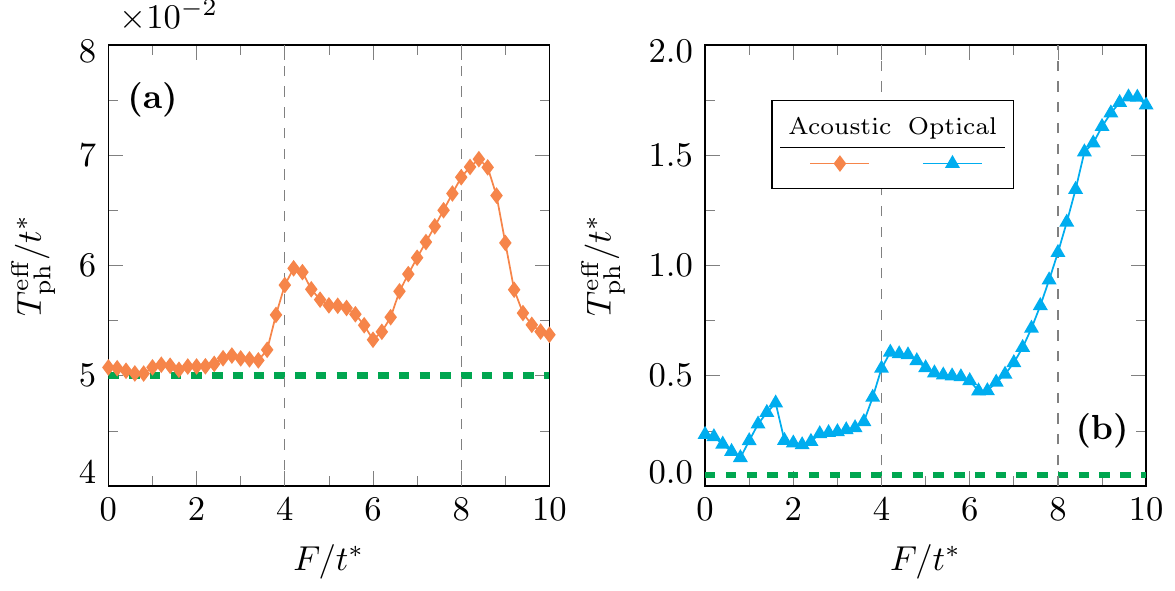}
 \caption{\resub{(a) The self-consistent acoustic phonon effective temperature $T^{\text{eff}}_{\text{ph}}$ obtained by minimization of the {\em cost function}~\eqref{eq:ph_cost_function} is shown as function of the applied field $F$. (b) Same for the self-consistent optical phonon. Dashed vertical black lines mark the resonances $F=U/2$ and $F=U$, while dashed horizontal green lines mark the acoustic and optical phonon effective temperature in the non-self-consistent case (again obtained according to Eq.~\eqref{eq:ph_cost_function}), which equals the equilibrium temperature of the baths $T=0.05t^{\ast}$. Default parameters for both cases can be found in Tab.~\ref{tab:default_pars}. (Here $\Gamma_{\text{e}}=0.12t^{\ast}$ and $U=8t^{\ast}$.)}}
 \label{fig:effective_Temp_comparison}
\end{figure}

\resub{The results are shown in Fig.~\ref{fig:effective_Temp_comparison}. In order to check the validity of Eq.~\eqref{eq:ph_cost_function}, we first use it to extract the phonon effective temperature of the NSC case, for both acoustic and Holstein phonons. Notably, in the NSC case our method yields the exact equilibrium temperature of the bath $T^{\text{eff}}_{\text{ph}}=0.05t^{\ast}$ (corresponding to $\beta=20/t^{\ast}$ as reported in Tab.~\ref{tab:default_pars}), see the green horizontal dashed lines in Figs~\ref{fig:effective_Temp_comparison}(a) and (b).

On the other hand, when considering the SC case, the acoustic phonon effective temperature $T^{\text{eff}}_{\text{ph}}$ shows the two-peak structure which is typical of the current $J$ (see Fig.~\ref{fig:Obs_ac_ph} for instance) with main resonances at $F\approx U/2$ and $F\approx U$, Fig.~\ref{fig:effective_Temp_comparison}(a). Notably, in this setup the highest $T^{\text{eff}}_{\text{ph}}$ reached by the acoustic phonons is still of the same order of magnitude of the equilibrium temperature of the bath $T=0.05t^{\ast}$ (notice the scale). On the other hand, in the case of SC optical phonons, Fig.~\ref{fig:effective_Temp_comparison}(b), the effective temperature $T^{\text{eff}}_{\text{ph}}$ still exhibits the characteristic two-peak structure at the main resonances, but the one at $F\approx U$ is shifted towards values of $F$ where the current $J$ is already suppressed (see Fig.~\ref{fig:observables_Hols}). Also, optical phonons undergo a much higher heating process, as can be inferred by the scale in Fig.~\ref{fig:effective_Temp_comparison}(b).
We then conclude that acoustic phonons are much more effective in dissipating the current-induced Joule heat, thus affecting the conducting properties just slightly, as compared to optical phonons. The reason is that, at self-consistency, the former undergo a negligible heating process with respect to the latter as shown in Fig.~\ref{fig:effective_Temp_comparison}.
}

\section{Conclusions}\label{sec:conclusions}
In this work we study the response of a Mott-insulating system to a constant electric field. We focus on a single-band Hubbard model attached to fermionic baths, with the inclusion of either optical or acoustic phonons as dissipation mechanism. The introduction of phonons is crucial for a correct description of heat transfer within the system upon approaching the \resub{current-conducting state} and thus of the dielectric breakdown. We show that by employing optical phonons within a self-consistent scheme, the steady-state current is sensibly suppressed with respect to the nonself-consistent case \resub{when the field equals the band gap}. This reduction of the current is accompanied by an increase in the phonon temperature, signalling the exchange of heat between phonons and the hot electrons of the lattice. Also, we find that the temperature of the baths does not affect these results, as the latter is smaller than the phonons characteristic frequency.

On the other hand, in the case of acoustic phonons, self-consistency does not influence the current characteristics significantly. Its effect can be detected in a slight reduction of the steady-state current at field strengths close to half of the gap, and thus away from the \resub{current-conducting state}, especially for very small values of the phonon cutoff frequency. This seems to confirm that long-wavelength (acoustic) phonons are well-suited to dissipate the excess heat, \resub{as suggested by the analysis of the phonon effective temperature at the end of Sec.~\ref{sec:temp_dep_ac_ph}}. Also, in contrast to optical phonons the steady-state current seems to be slightly dependent on the temperature in this case. 
This aspect is most likely related to the fact that acoustic phonons can have very small energies in their spectrum so they are prone to be affected by any arbitrarily low temperature. 

\begin{acknowledgments}
We thank J. Lotze for contributing theoretical discussions and useful insights. This research was funded by the Austrian Science Fund (Grant No. P 33165-N) and by NaWi Graz. The results have been obtained using the Vienna Scientific Cluster and the D-Cluster Graz.
\end{acknowledgments}

\begin{center}
{\bf Author information}
\end{center}

{\bf Contributions:} E.A. conceived the project and supervised the work. Code development: T.M.M. contributed phonons implementation and produced theoretical data, D.W. contributed configuration interaction impurity solver, P.G. helped in developing the phonon implementation and E.A. contributed improvements in the fitting routine. The Manuscript was drafted by T.M.M. with contributions from all authors.

\appendix

\section{Floquet Green's functions formalism}\label{sec:GFs_Dyson_Floquet}

Under the action of a static field the system is time-translational invariant~\cite{ma.ga.22,ts.ok.08} and so is the corresponding GF in the Coulomb gauge. Within the temporal gauge, the GFs will fulfil the periodicity condition $X(t,t^{\prime})=X(t+\tau,t^{\prime}+\tau)$ with $\tau=2\pi/\Omega$ being the period and $X$ denoting any component of the {\em Keldysh} GF. According to the so-called \emph{Keldysh-Floquet} formalism~\cite{ts.ok.08,sc.mo.02u,jo.fr.08}, the GF can be represented as (for simplicity we drop the crystal momentum $\vec{k}$)
\begin{equation}\label{eq:FloquetGF}
\kel{X}_{mn}(\omega) =\int \dd t_{\text{rel}} \int_{-\tau/2}^{\tau/2} \frac{\dd t_{\text{av}}}{\tau} \ee^{\ii[\left(\omega+m\Omega\right) t -\left( \omega+n\Omega\right)t^{\prime}]} \kel{X}(t,t^{\prime}),
\end{equation}
where $t_{\text{rel}} = t-t^{\prime}$ and $t_{\text{av}} = (t+t^{\prime})/2$ are the relative and average time variables. It should be noted that any Floquet-represented quantity~\eqref{eq:FloquetGF} can be recast as
\begin{equation}\label{eq:WignerGF}
\kel{X}_{l}(\omega^{\prime}) =\int \dd t_{\text{rel}} \int_{-\tau/2}^{\tau/2} \frac{\dd t_{\text{av}}}{\tau} \ee^{\ii l\Omega t_{\text{av}} + \ii \omega^{\prime} t_{\text{rel}}} \kel{X}(t,t^{\prime}),
\end{equation}
which is usually referred to as the {\em Wigner representation}~\cite{ts.ok.08,ma.ga.22} and can be easily derived from~\eqref{eq:FloquetGF} with
\begin{align}\label{eq:Fl_2_Wig_mapping}
\begin{split}
\omega^{\prime} & =\omega+(m+n)\Omega/2 \\ 
l & =m-n.
\end{split}
\end{align}
It is also worth recalling the shifting property of any Floquet GF
\begin{equation}\label{eq:Fl_shift}
\kel{X}_{mn}(\omega) = \kel{X}_{m-n,0}(\omega+n\Omega),
\end{equation}
which can be easily derived from~\eqref{eq:FloquetGF}.

\section{Floquet-DMFT and discussion on the accuracy of the AMEA impurity solver}\label{sec:imp_solver}

\subsection{Floquet-dynamical mean-field theory}

The aim of this section is to briefly review the key aspects of the F-DMFT. For further details we refer to the previous works~\cite{ts.ok.08,so.do.18,ma.ga.22}.

Within the F-DMFT, the interacting problem~\eqref{eq:FullDysonEq} is solved by neglecting non-local contributions to the electron SE, i.e. $\kel{\mat{\Sigma}}(\omega,\epsilon,\overline{\epsilon}) \to \kel{\mat{\Sigma}}(\omega)$, and the original problem is mapped onto a single-site impurity model coupled to a bath hybridization function $\kel{\mat\Delta}(\omega)$ acting like a {\em reservoir} on a mean field level. The impurity GF $\kel{\mat{G}}_{\text{imp}}$ obeys the following equation
\begin{equation}\label{eq:imp_Dyson_eq}
\kel{\mat{G}}^{-1}_{\text{imp}}(\omega) = \kel{\mat{g}}^{-1}_{0,\text{site}}(\omega) - \kel{\mat{\Delta}}(\omega) - \kel{\mat{\Sigma}}(\omega),
\end{equation}
with the single site {\em retarded} GF $[g_{0,\text{site}}^{-1}]^{\text{R}}=\omega-\varepsilon_{\text{c}}$. The hybridization $\kel{\mat\Delta}(\omega)$ is determined from~\eqref{eq:imp_Dyson_eq} requiring $\kel{\mat{G}}_{\text{imp}}(\omega)\overset{!}{=}\kel{\mat{G}}_{\text{loc}}(\omega)$ at self-consistency. Notice that the gauge-invariant GF in the case of a static field is diagonal in Floquet indices~\cite{ts.ok.08}, thus by employing the translation symmetry $\kel{X}_{mm}(\omega) = \kel{X}_{00}(\omega+m\Omega)$, see Eq.~\eqref{eq:Fl_shift}, one restricts the problem to the computation of the $(0,0)$ Floquet matrix-element alone for all the quantities in~\eqref{eq:imp_Dyson_eq}. 

\begin{figure}[h]
\includegraphics[width=\linewidth]{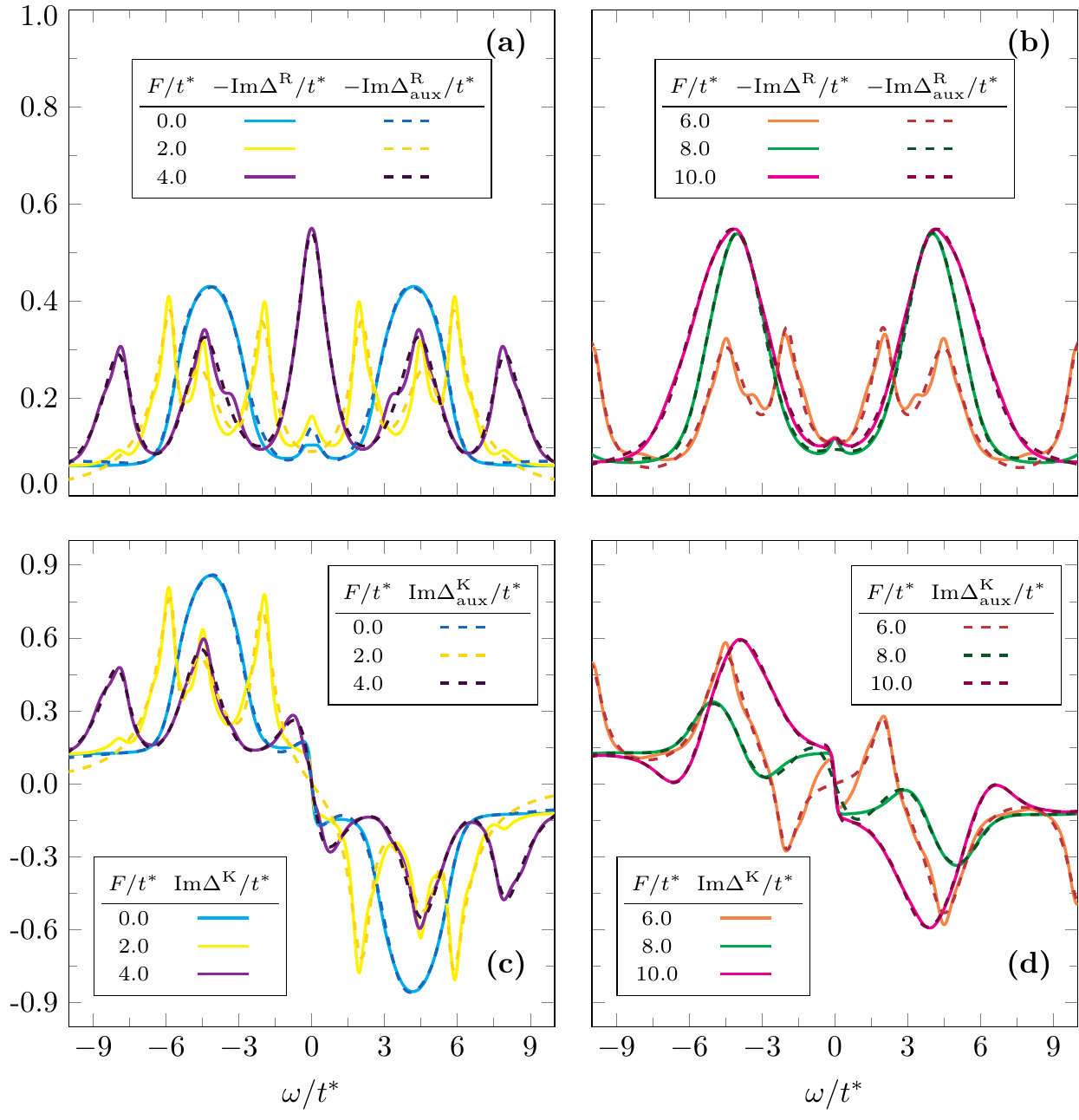}
\caption{SC phonons: panel (a) shows the imaginary parts of $\Delta^{\text{R}}$ and $\Delta^{\text{R}}_{\text{aux}}$ for the applied field $F/t^{\ast}=\left\{ 0, 2, 4 \right\}$. In panel (b) are shown the same quantities for $F/t^{\ast}=\left\{ 6, 8, 10 \right\}$. The imaginary parts of $\Delta^{\text{K}}$ and $\Delta^{\text{K}}_{\text{aux}}$ are shown for (c) $F/t^{\ast}=\left\{ 0, 2, 4 \right\}$ and (d) $F/t^{\ast}=\left\{ 6, 8, 10 \right\}$. The number of bath sites is $N_{\text{B}}=6$. Default parameters are specified in setup O of Tab.~\ref{tab:default_pars}. (\resub{Here $\Gamma_{\text{e}}=0.12t^{\ast}$ and $U=8t^{\ast}$.})}
\label{fig:hybs_fit_Ein_ph_NB6}
\end{figure}

\subsection{Accuracy of the AMEA impurity solver}

In order to get the electron SE required in Eq.~\eqref{eq:imp_Dyson_eq} we employ the so-called auxiliary master equation approach (AMEA)~\cite{ar.kn.13,do.nu.14,do.ga.15,do.so.17,ar.do.18}, a non-equilibrium impurity solver which relies on an \emph{auxiliary} open quantum system consisting of a finite number of bath sites $N_{\text{B}}$ attached to Markovian reservoirs obeying the Lindblad equation. The optimal bath parameters are obtained by fitting
the non-interacting hybridization function $\kel{\Delta}_{\text{aux}}$ of this auxiliary system to $\kel{\Delta}$ from the DMFT iteration.
The accuracy of the impurity solver is then directly related to the difference between
$\kel{\Delta}_{\text{aux}}$ and $\kel{\Delta}$, which decreases exponentially with increasing $N_{\text{B}}$~\cite{do.so.17}. 
The auxiliary Lindblad problem is solved by many-body exact-diagonalization (ED) techniques for open quantum systems.

\begin{figure}[t]
\includegraphics[width=\linewidth]{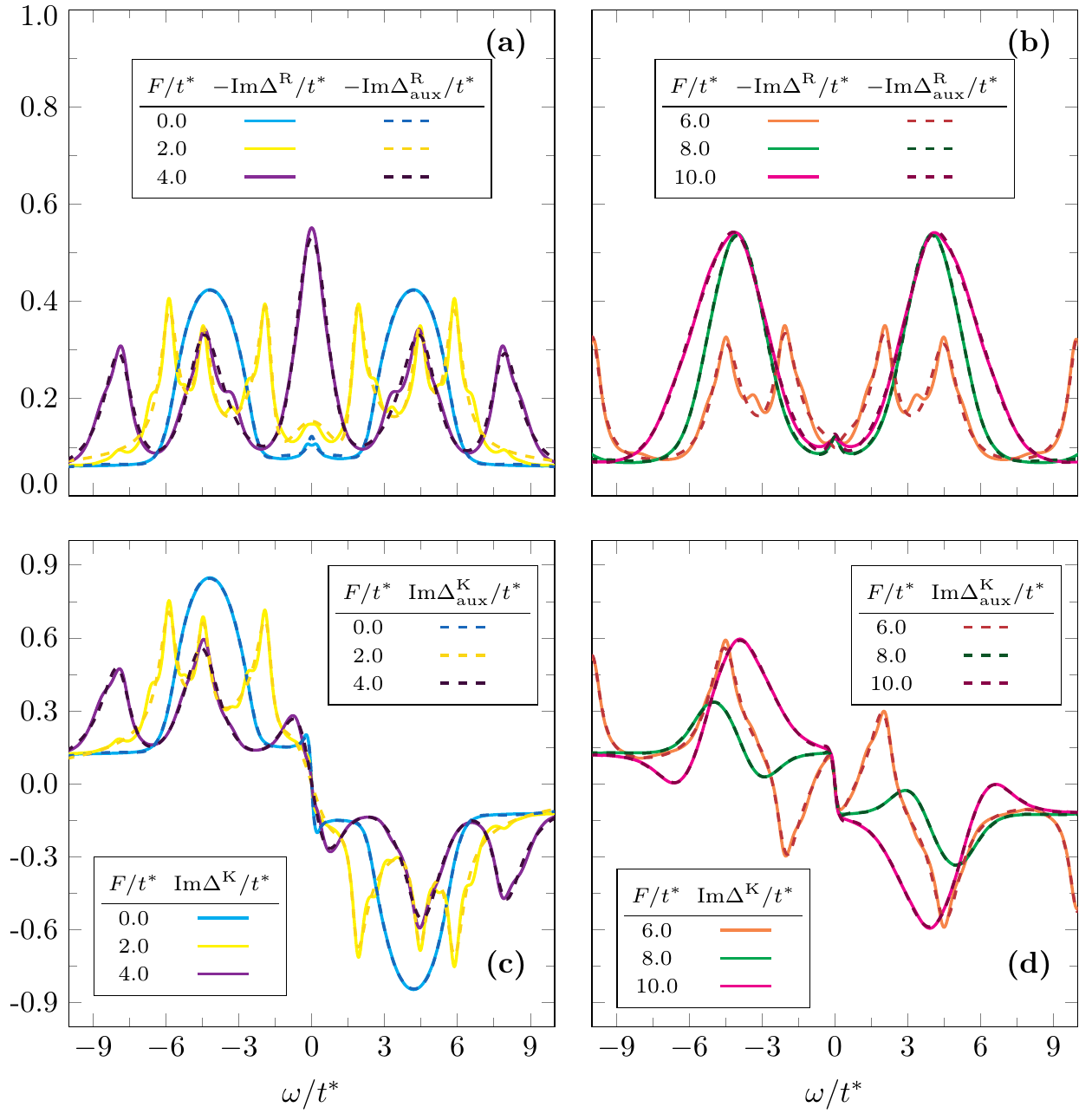}
\caption{Same as Fig.~\ref{fig:hybs_fit_Ein_ph_NB6} with $N_{\text{B}}=8$. Default parameters are specified in setup O of Tab.~\ref{tab:default_pars}. (\resub{Here $\Gamma_{\text{e}}=0.12t^{\ast}$ and $U=8t^{\ast}$.)}}
\label{fig:hybs_fit_Ein_ph_NB8}
\end{figure}

\begin{figure}[b]
\includegraphics[width=\linewidth]{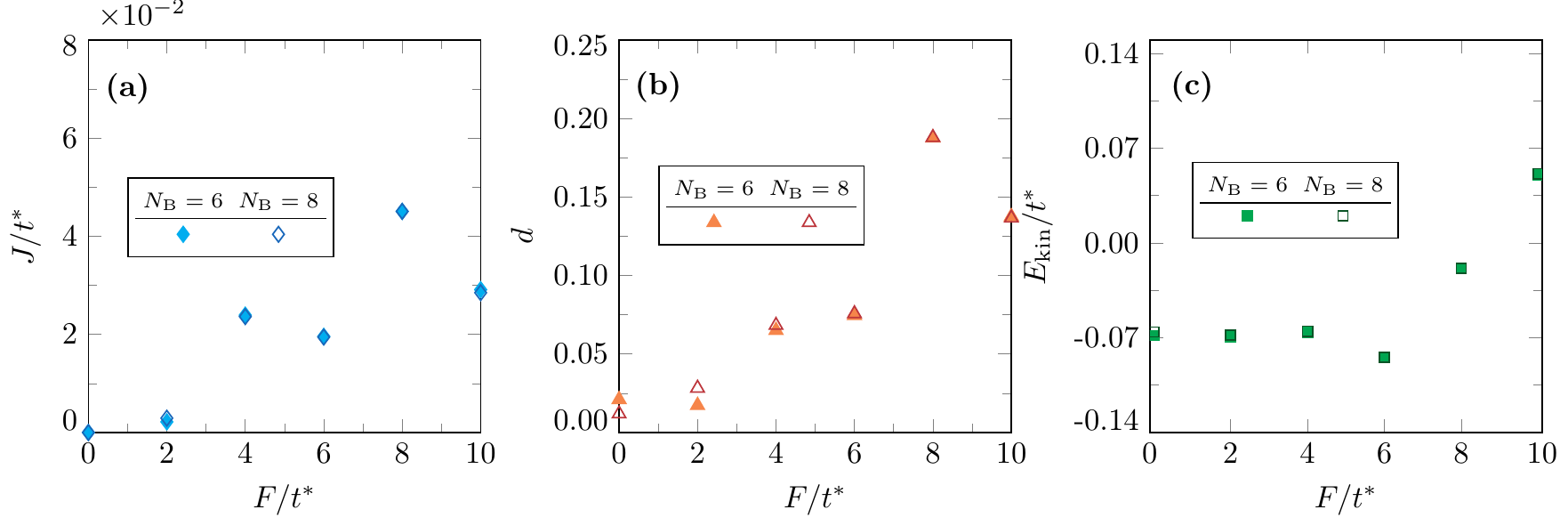}
\caption{Difference in the observables as obtained with $N_{\text{B}}=6$ and $N_{\text{B}}=8$ for the (a) current $J$, (b) double occupation $d$ and (c) kinetic energy $E_{\text{kin}}$. Default parameters are specified in setup O in Tab.~\ref{tab:default_pars}. (\resub{Here $\Gamma_{\text{e}}=0.12t^{\ast}$ and $U=8t^{\ast}$.})}
\label{fig:obs_diff_Ein_ph_NB6_8}
\end{figure}

Since for a given $N_{\text{B}}$ the number of fit parameters~\cite{do.so.17} is roughly four times the number of parameters one would have that for a conventional equilibrium ED impurity solver~\cite{ge.ko.96}, a relatively small $N_{\text{B}}\sim 4-8$ is generally sufficient for a reasonable convergence depending on the situation~\cite{ti.do.15,ti.do.16}.
In order to illustrate this, in
 this section we compare the accuracy of the results obtained from the AMEA impurity solver obtained with a two different values of $N_{\text{B}}=\left\{ 6,8\right\}$.
In particular, $N_{\text{B}}=8$ was only possible within ED~\footnote{Notice, that for a given number of lattice sites, the dimension of the many-body ``super'' Hilbert space for the open quantum system is the square of the dimension of a corresponding closed system.}
 thanks to a recently implemented~\cite{we.lo.22u}
 configuration interaction~\cite{zg.gu.12,li.de.13} (CI) approach to this open system many-body problem. A detailed description can be found in Ref.~\cite{we.lo.22u}.
One should add that
 the CI approach also
 drastically reduces the computational time required to solve the impurity problem 
in comparison 
to conventional Krylov-space methods~\cite{do.nu.14} for the same number of bath sites $N_{\text{B}}=6$. 
Since the $N_{\text{B}}=8$ are numerically expensive, we only carry out our benchmark for selected  values of the electric field $F$.

We start by comparing the imaginary parts of the bath hybridization functions $\kel{\Delta}$ and $\kel{\Delta}_{\text{aux}}$ corresponding to exemplary values of the applied field $F$ for $N_{\text{B}}=\left\{ 6, 8 \right\}$.
Data concerning $N_{\text{B}}=6$ can be found in Fig.~\ref{fig:hybs_fit_Ein_ph_NB6}, where we observe that $\text{Im}\Delta^{\text{R}}_{\text{aux}}$ either overestimates ($F=0$) or underestimates ($F=2t^{\ast}$) the peak at $\omega\approx 0$, see panel (a). This low-frequency peak shows up also at $F=8t^{\ast}$ and it is not captured in this situation either, see panel (b). In panel (c) we see that the inflection point in the low-frequency range is not well reproduced by $\text{Im}\Delta^{\text{K}}_{\text{aux}}$ especially at $F=2t^{\ast}$. The same thing happens for $F=6t^{\ast}$, see panel (d).

\begin{figure}[t]
\includegraphics[width=\linewidth]{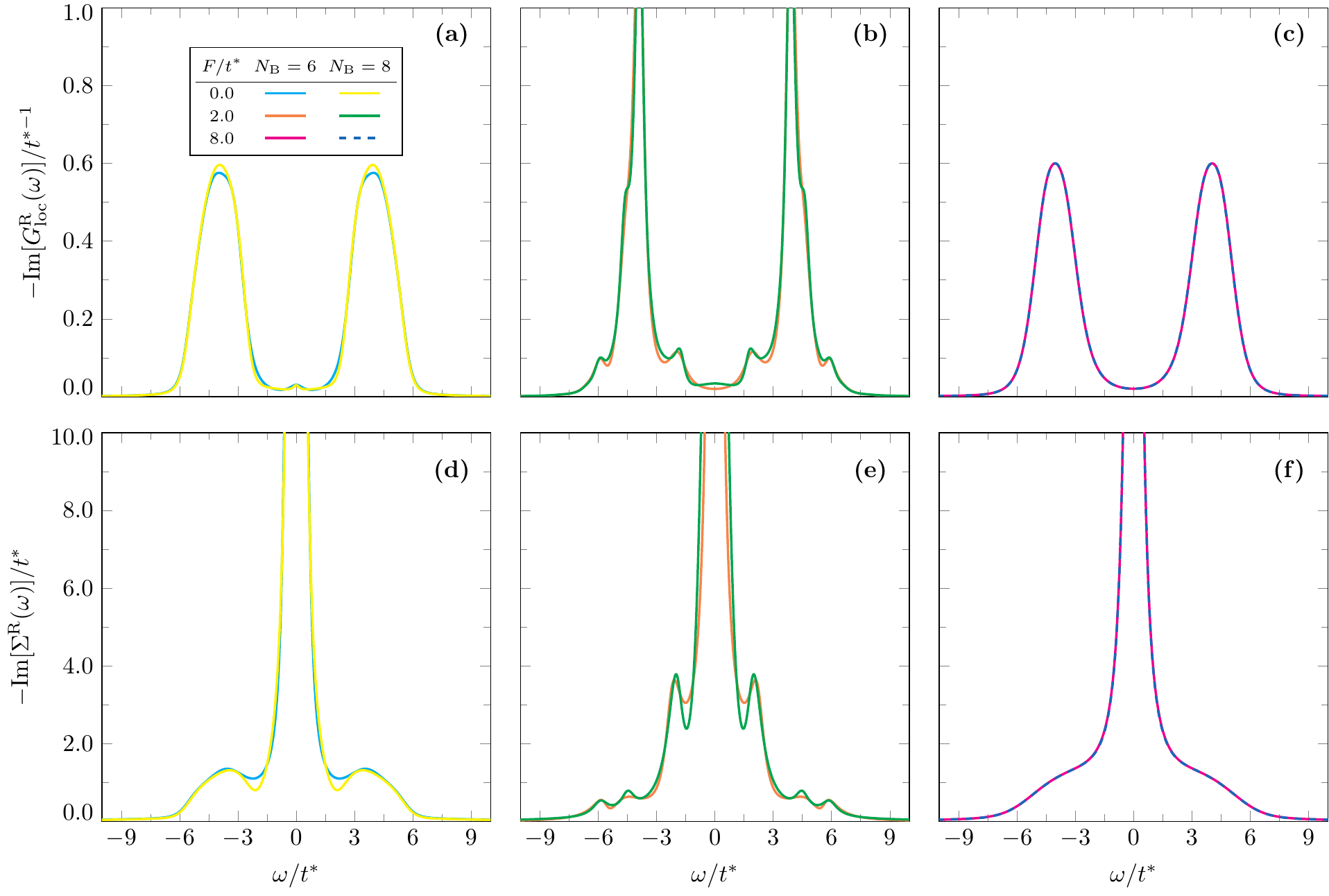}
\caption{SC phonons. First row: imaginary part of the retarded electron GF $\text{Im}[G^{\text{R}}_{\text{loc}}(\omega)]$ for selected field strengths obtained with $N_{\text{B}}=6$ and $N_{\text{B}}=8$. Second row: the imaginary part of the retarded electron SE $\text{Im}[\Sigma^{\text{R}}(\omega)]$ for the same situations. Default parameters are specified in setup O of Tab.~\ref{tab:default_pars}. (\resub{Here $\Gamma_{\text{e}}=0.12t^{\ast}$ and $U=8t^{\ast}$.)}}
\label{fig:Ret_GFs_SEs_Ein_ph_NB}
\end{figure}

\begin{figure}[t]
\includegraphics[width=\linewidth]{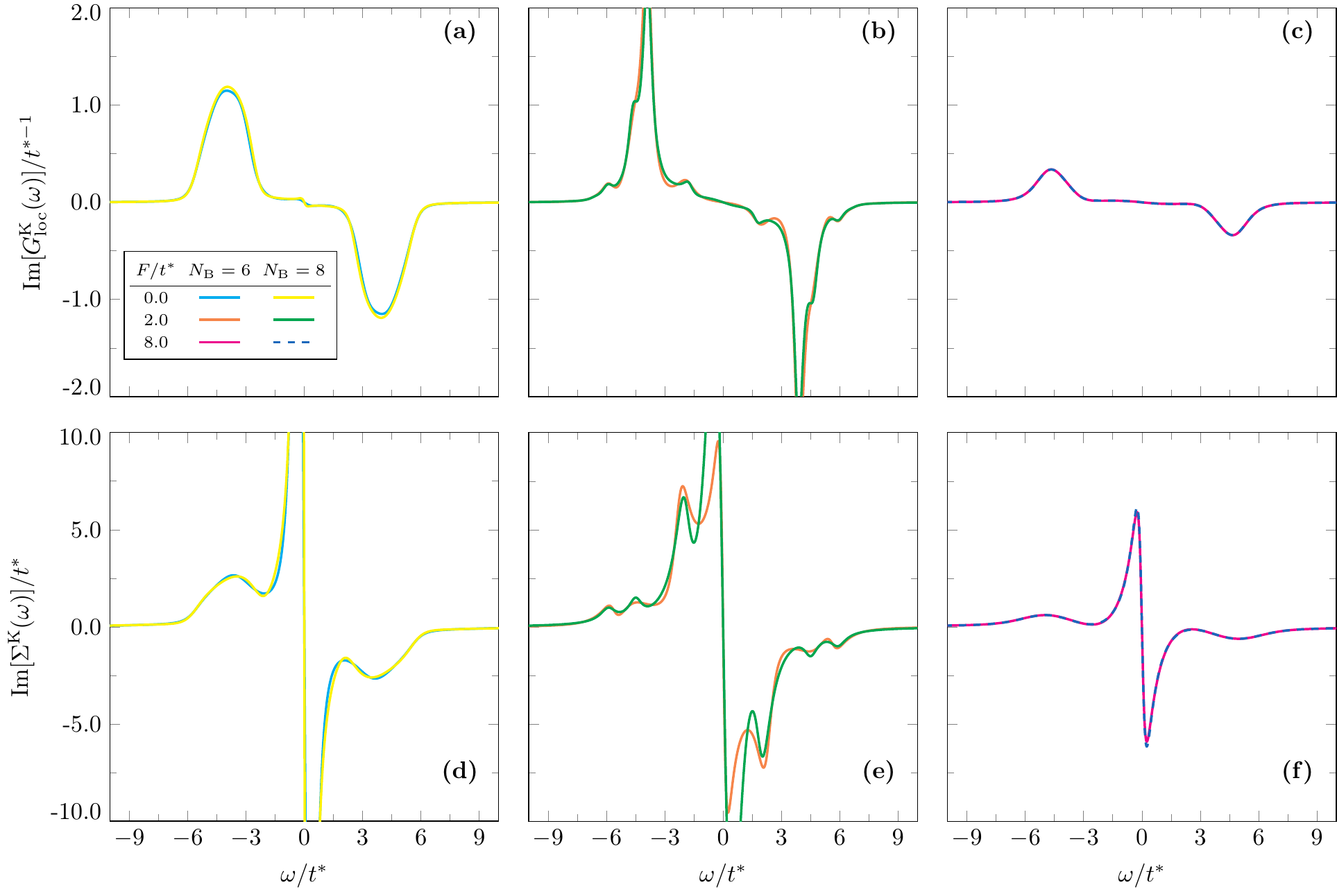}
\caption{SC phonons. First row: imaginary part of the Keldysh electron GF $\text{Im}[G^{\text{K}}_{\text{loc}}(\omega)]$ for selected field strengths obtained with $N_{\text{B}}=6$ and $N_{\text{B}}=8$. Second row: the imaginary part of the Keldysh electron SE $\text{Im}[\Sigma^{\text{K}}(\omega)]$ for the same situations. Default parameters are specified in setup O of Tab.~\ref{tab:default_pars}. (\resub{Here $\Gamma_{\text{e}}=0.12t^{\ast}$ and $U=8t^{\ast}$.)}}
\label{fig:Kel_GFs_SEs_Ein_ph_NB}
\end{figure}

On the other hand, with $N_{\text{B}}=8$ the peak at $\omega\approx 0$ is quite well reproduced by $\text{Im}\Delta^{\text{R}}_{\text{aux}}$ for $F=0$, $F=2t^{\ast}$, see Fig.~\ref{fig:hybs_fit_Ein_ph_NB8}(a), and for $F=8t^{\ast}$, see panel (b). Also, the inflection points in the low-frequency range of $\text{Im}\Delta^{\text{K}}_{\text{aux}}$ at $F=2t^{\ast}$ and $F=6t^{\ast}$ are now perfectly distinguished, see panels (c) and (d).

Of course, the relevant question is, how much the observables change by increasing the number of bath sites. In Fig.~\ref{fig:obs_diff_Ein_ph_NB6_8} we compare the current, double occupation and kinetic energy obtained with $N_{\text{B}}=6$ and $N_{\text{B}}=8$. As we can see from panels (a) to (c) one hardly sees any difference in $J$, $d$ and $E_{\text{kin}}$ between the results obtained with the two values of $N_{\text{B}}$. This shows that these results are converged with respect to $N_{\text{B}}$.

A {\em slight} difference can be seen when comparing the imaginary parts of the retarded and Keldysh electron GFs and corresponding SEs.

In Fig.~\ref{fig:Ret_GFs_SEs_Ein_ph_NB} we show the differences in $\text{Im}[G^{\text{R}}_{\text{loc}}(\omega)]$ and $\text{Im}[\Sigma^{\text{R}}(\omega)]$. Panels (a), (b), (d) and (e) show that at $F=0$ and $F=2t^{\ast}$ the profiles of both the GF and SE only slightly change around specific $\omega$ regions for the two different
$N_{\text{B}}$ values. Panels (c) and (f), instead, show the perfect matching of the GFs and the SEs at the \resub{current conducting state} $F=8t^{\ast}$.

Finally, Fig.~\ref{fig:Kel_GFs_SEs_Ein_ph_NB} compares the imaginary part of the {\em Keldysh} electron GFs and SEs obtained by different values of $N_{\text{B}}$. Again, slight differences can be seen in panels (a), (b), (d) and (e) corresponding to  $F=0$ and $F=2t^{\ast}$, while for $F=8t^{\ast}$ (panels (c) and (f)), no differences can be seen.

\section{Real-time Keldysh components of $\Sigma_{\text{e-ph}}$ and $\Pi_{\text{e-ph}}$}\label{sec:real-time_eph_se}

\begin{figure}[b]
\includegraphics[width=\linewidth]{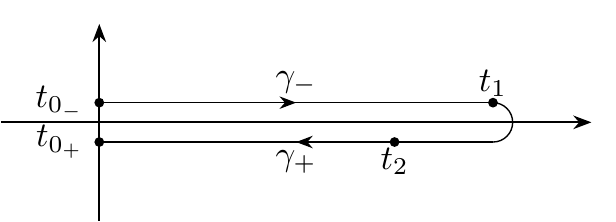}
\caption{Keldysh contour $C_{\kappa}=\gamma_{-} \cup \gamma_{+}$ for real-time arguments.}
\label{fig:keldysh_contour}
\end{figure}

The Keldysh components of the e-ph SE in Eq.~\eqref{eq:backbone_e-ph_SE} are obtained by means of the \emph{Langreth rules} \cite{st.va.13} and read
\begin{align}\label{eq:TD_e-ph_SE}
\begin{split}
\Sigma^{\text{R}}_{\text{e-ph}}(t,t^{\prime})& = \ii g^{2} \left( G^{\text{R}}(t,t^{\prime})D^{>}_{\text{ph}}(t,t^{\prime}) + G^{<}(t,t^{\prime})D^{\text{R}}_{\text{ph}}(t,t^{\prime}) \right), \\
\Sigma^{\text{K}}_{\text{e-ph}}(t,t^{\prime}) & = \ii g^{2} \left[ G^{\text{K}}(t,t^{\prime})D^{>}_{\text{ph}}(t,t^{\prime}) \ + \right. \\ 
& \left. + \ G^{<}(t,t^{\prime}) \left(D^{>}_{\text{ph}}(t,t^{\prime})-D^{<}_{\text{ph}}(t,t^{\prime})\right) \right], \\
\end{split}
\end{align}
where $t,t^{\prime}$ lie on the Keldysh contour $C_{\kappa}$ shown in Fig.~\ref{fig:keldysh_contour}.

Analogously, we derive the real time components of the polarization diagram in Eq.~\eqref{eq:bubble_GG}
\begin{align}\label{eq:TD_e-ph_bubble}
\begin{split}
\Pi^{\text{R}}_{\text{e-ph}}(t,t^{\prime}) & = -2\ii g^{2} \left( G^{\text{R}}(t,t^{\prime}) G^{<}(t^{\prime},t) + G^{<}(t,t^{\prime}) G^{\text{A}}(t^{\prime},t)\right), \\
\Pi^{\text{K}}_{\text{e-ph}}(t,t^{\prime}) & = -2\ii g^{2} \left( G^{>}(t,t^{\prime}) G^{<}(t^{\prime},t) + G^{<}(t,t^{\prime}) G^{>}(t^{\prime},t)\right),
\end{split}
\end{align}
and recall the relations $G^{\text{A}}(t,t^{\prime}) = [ G^{\text{R}}(t^{\prime},t) ]^{\ast}$ and $G^{\lessgtr}(t,t^{\prime})=-[G^{\lessgtr}(t^{\prime},t)]^{\ast}$.

\bibliographystyle{prsty}
\bibliography{references_database,my_refs}

\end{document}